\documentclass[bibyear]{aa}

\usepackage{graphicx}
\usepackage{txfonts}
\usepackage[breaklinks=true]{hyperref}
\hypersetup{
colorlinks = true,
citecolor = black,
linkcolor=blue,
urlcolor = blue
}

\usepackage{natbib, twoopt}
\usepackage{textcomp, gensymb}
\usepackage{multirow}
\usepackage{longtable}
\usepackage{supertabular}
\usepackage{amsmath}
\usepackage{amssymb}
\usepackage[T1]{fontenc}

\usepackage{booktabs}
\usepackage{tabularx}
\usepackage{array}

\makeatletter
\renewcommand*\aa@pageof{, page \thepage{} of \pageref*{LastPage}}
\makeatother

\begin{document}

   \title{Variable stars in the residual light curves of OGLE-IV eclipsing binaries towards the Galactic Bulge}
   
   \titlerunning{Background variables in OGLE-IV eclipsing binaries}

   \author{R. Z. \'{A}d\'{a}m \inst{1}\fnmsep\inst{2}
          \and
          T. Hajdu\inst{1}\fnmsep\inst{2}\fnmsep\inst{3}\fnmsep\inst{6}
          \and
          A. Bódi\inst{2}\fnmsep\inst{3}
          \and
          R. Hajdu\inst{4}
          \and
          T. Szklenár\inst{2}\fnmsep\inst{3}
          \and
          L. Molnár\inst{2}\fnmsep\inst{3}\fnmsep\inst{5}
          }

   \institute{E\"{o}tv\"{o}s Lor\'{a}nd University, Department of Astronomy, H-1117 P\'{a}zm\'{a}ny P\'{e}ter s\'{e}t\'{a}ny 1/A, Budapest, Hungary\\
              \email{adam.rozalia@csfk.org}
         \and
             Konkoly Observatory, Research Centre for Astronomy and Earth Sciences, ELKH, MTA Centre of Excellence, Konkoly Thege Mikl\'os \'ut 15-17, H-1121 Budapest, Hungary
        \and
            MTA CSFK Lend\"ulet Near-Field Cosmology Research Group
        \and
            \'{O}buda University, Department of Computer Engineering, H-1034 B\'{e}csi \'{u}t 96/b, Budapest, Hungary
        \and
            E\"otv\"os Lor\'and University, Institute of Physics, H-1117 P\'azm\'any P\'eter s\'et\'any 1/A, Budapest, Hungary
        \and
            Eszterh\'{a}zy K\'{a}roly Catholic University, Department of Physics, H-3300 Eszterh\'{a}zy t\'{e}r 1, Eger, Hungary 
             }

   \date{Received XXX; accepted YYY}

  \abstract
  {The Optical Gravitational Lensing Experiment (OGLE) observed around 450,000 eclipsing binaries (EBs) towards the Galactic Bulge. Decade-long photometric observations such as these provide an exceptional opportunity to thoroughly examine the targets. However, observing dense stellar fields such as the Bulge may result in blends and contamination by close objects.}
  {We searched for periodic variations in the residual light curves of EBs in OGLE-IV and created a new catalogue for the EBs that contain `background' signals after the investigation of the source of the signal.}
  {From the about half a million EB systems, we selected those that contain more than 4000 data points. We fitted the EB signal with a simple model and subtracted it. To identify periodical signals in the residuals, we used a GPU-based phase dispersion minimisation python algorithm called \texttt{cuvarbase} and validated the found periods with Lomb-Scargle periodograms. We tested the reliability of our method with artificial light curves.}
  {We identified 354 systems where short-period background variation was significant. In these cases, we determined whether it is a new variable or just the result of contamination by an already catalogued nearby one. We classified 292 newly found variables into EB, $\delta$~Scuti, or RR~Lyrae categories, or their sub-classes, and collected them in a catalogue. We also discovered four new doubly eclipsing systems and one eclipsing multiple system with a $\delta$~Scuti variable, and modelled the outer orbits of the components.} 
    {}


   \maketitle
 
\section{Introduction}

Eclipsing binary (EB) stars belong to the large family of multiple stellar systems. They are close binaries whose light curves (LCs) show periodic brightness variations due to the motion and mutual occultations of the components in our line of sight. 
Their investigation has great astrophysical importance, not only because the fundamental parameters of the components, such as masses, radii, and temperatures, are uniquely determinable \citep{Andersen1991,Wilson1971,Wilson2012,Prsa2005,Prsa2016}, but also because their eclipse timing variation (ETV) analysis provides us with information about the multiplicity of the systems \citep{Borkovits_2015,Zasche2016,Zasche2017, Hajdu2019, Hajdu2022}. These systems can also be used to test and develop stellar evolution models (e.g. \citealt{2014Reipurth}). 
As \cite{2021Tokovinin} summarises, all hierarchical systems can be traced back to nested binaries, and thus their structure and evolution can be approximated as a set of stellar pairs. Nevertheless, studying isolated binaries is not always the key to understanding high-order systems, since there are evolutionary pathways that are only apparent in the evolution of triples \citep{2020Toonen,2022Toonen}.
Multiple star systems have various evolutionary paths, many of which result in some component temporarily becoming a variable star.

The detection of a pulsating star in an EB in particular can be very valuable, as it allows us to study the evolution and properties of pulsating stars in more detail. It also makes it possible to test the accuracy of pulsation models against model-independent measurements.

Searching for Cepheids in binary systems, for example, has been at the centre of attention for decades now.
The realisation that the dynamical masses and the masses calculated from evolutionary and pulsation models do not align (also known as the Cepheid mass problem) dates back to the 1960s (\citealt{Cox_1980} and references therein). Over the last five decades there have been many attempts to resolve this conundrum (e.g. \citealt{Moskalik_1992, Neilson_2011}). A major breakthrough came with OGLE-LMC-CEP-0227, which is a Cepheid residing in an EB. \cite{P_Nature_2010} were able to calculate its mass with 1\% accuracy, which led to the inspection and correction of evolution models (e.g. \citealt{Cassisi_2011,PradaMoroni_2012}).
Since then, the masses of other Cepheids have been determined \citep{Pietr_2011,Gieren_2014,Pilecki_2015}, in all cases for Cepheids that are in EB systems. Hence it is worth looking for this variable type the other way around, by studying EBs.

RR Lyrae stars make up a large portion of variable stars, and they are cosmic structure tracers and thoroughly studied objects.Yet, there is still no known confirmed binary system with an RR Lyrae star. OGLE-BLG-RRLYR-02792 had recently been classified as the first known RR Lyrae in an EB, but further analysis found its mass to be considerably lower than masses predicted for RR Lyrae variables (\citealt{Pietrzynski_2012}). The system was instead re-classified as a binary evolution pulsator (BEP). \cite{Karczmarek_2017} studied the occurrence and abundance of this type and stated that 0.8\% of RR Lyrae stars might be misclassified BEPs. Moreover, in a recent study \cite{Bobrick_arxiv_2022} presented and discussed how young and metal-rich RR Lyrae stars might indeed be the result of binary interactions, suggesting that wide binaries must exist among them. We note that even if there is no confirmed RR Lyrae binary yet, there are a few dozen suitable candidates from pulsation $O-C$ studies that require follow-up spectroscopic observations (\citealt{Hajdu_2015,Hajdu_2021}).

A considerable fraction of $\delta$ Scuti stars are members of binary systems \citep{Liakos_Niarchos_2017}, providing an important opportunity to test stellar structure and evolution theory. In particular, such systems enable the determination of the current evolutionary status of components and the system age (see e.g. \citealt{Guo_2016,Streamer_2018}) and provide the possibility to study tidal interactions \citep{Handler_2020Nat}, the effect of the mass exchange, and its impact on the evolution of the system (see e.g. \citealt{Liakos_2022}).

Studying $\delta$ Scuti stars is also a valuable tool when searching for binary systems, as we can measure the effect of binary motion on stellar pulsations.
\cite{Shibahashi_Kurtz_2012} showed that high-precision LCs of pulsating stars provide a way to derive radial velocity curves solely from photometric data. \citet{Murphy_2014} developed a complementary method by using phase modulation instead of frequency modulation, which is more suitable for wider binaries and more sensitive to variations.

The Optical Gravitational Lensing Experiment \citep[OGLE;][]{Ogle_tortenet0}, a long-running ground-based photometry project, observed around half a million EBs, providing the second largest EB catalogue for astronomers \citep{OGLE_BLG, OGLE_Magellanic} after the all-sky EB catalogue of \textit{Gaia} \citep{gaia-EB-2022}. The original goal of the OGLE project was to detect microlensing events caused by faint but massive compact objects, once considered to be significant contributors to dark matter. This necessitated the observation of areas with dense stellar backgrounds. However, because of the very crowded stellar fields, LCs of EB stars can be contaminated by close stars that cannot be resolved optically; some of these stars may even belong to the same system. Such contaminated systems were already found by the OGLE team\footnote{\url{http://www.astrouw.edu.pl/ogle/ogle4/OCVS/blg/ecl/remarks.txt}}, and the multiplicity of the listed Algol-type systems was tested by \citet{Zasche2019}. Nonetheless, there could be further systems that are not on this list. A comprehensive study is needed that includes not only the newly found variables, but also those where the presence of a nearby known star has a significant effect on the LC. A good example of this is the case of OGLE-BLG-ECL-157718 and OGLE-BLG-ECL-157729, where the frequency difference between the period of the binaries causes a significant ETV \citep{Hajdu2022}.

An investigation of the residual LCs of EBs would not only be important to get `clear' LCs, which is essential for modelling and understanding of the true nature of the systems, but also to find new pulsating variables, which can also be part of the EB. Furthermore, the LCs of EBs can be considered to be basically unchanging; therefore, subtracting the signal of the system has no effect on the signal of a possible blended variable.

The aim of this research is to identify systems whose LCs are significantly affected by other variables and to determine the origin of that signal. We attempt to determine whether the source belongs to the system or is just a background object. In Sect.~\ref{method} we present the details of the methods we used for our analyses. In Sect.~\ref{classification} we describe how we classified the newly found variables. Finally, in Sect.~\ref{result} we list the newly found variable stars and multiple systems. We summarise our results in Sect.~\ref{sec:summary}.

\section{Methods}
\label{method}
   
\begin{figure}[ht!]
    \centering
    \includegraphics[width=0.95\linewidth]{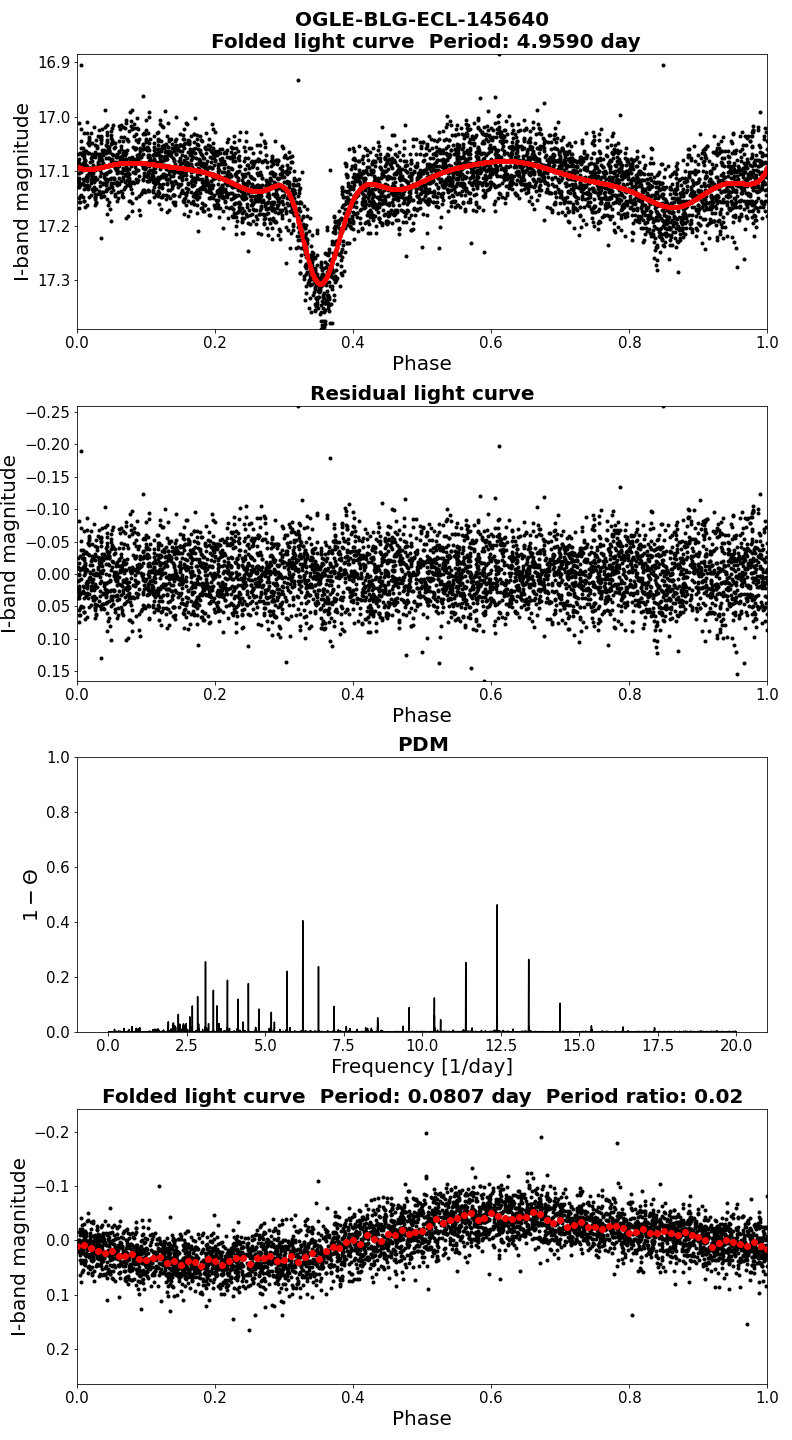}
    \caption{Workflow of the pulsating variable search. The top panel shows the folded LC of OGLE-BLG-ECL-145640 ($P_{\textup{EB}}=4.9590^d$) (black)
    and the EB trend derived from the \texttt{UnivariateSpline} fit (red). The second panel shows the residual LC, whose PDM is plotted in the third panel. In the bottom panel we present the new phase-folded LC of the residual folded by the period ($P_{\textup{new}}=0.0807^d$) found by the PDM analysis along with the folded and binned residual LC (red).
    }
    \label{fig:workflow}
\end{figure}

In recent years, many articles have been published on the identification of further variability in the LCs of EBs \citep[see e.g.][]{2019Gaulme, 2022Xinghao}, but these were primarily developed for the detection of pulsating variables. Here, we present an alternative way to search for variables in the background of EBs. Our goal was to identify as many periodic variable stars as possible with periods shorter than a few days ($P\lesssim10^d$). We expect most of them to be EBs instead of pulsating stars, based on their ubiquity and also on the number of objects in the various OGLE variable catalogues \citep{2018Soszynski_ogle_variables}.
In the following subsections we present the basic steps of this process, which are illustrated in Fig. \ref{fig:workflow}.

\subsection{System selection}
The OGLE Bulge EB catalogue contains around 450,000 systems and most of them have OGLE-IV measurements. For our investigation, similarly to \citet{Hajdu2019}, we used only those systems that were observed during phase IV of OGLE and whose \textit{I}-band LC contain more than 4000 data points. After these restrictions around 80,000 systems were analysed automatically with our custom-made program. 

\subsection{Subtraction of the EB signal} \label{sec:subtraction}
As a first step, the program produced the folded and binned light curve (FBLC) of the system using the catalogued period, for which the number of bins were based on the assumption presented in \cite{Bodi2021}. Next, we created a template function 
by fitting the FBLC with the \texttt{UnivariateSpline}\footnote{\href{https://docs.scipy.org/doc/scipy/reference/generated/scipy.interpolate.UnivariateSpline.html}{https://www.scipy.org/scipy.interpolate.UnivariateSpline.html}} function of the \texttt{Interpolate} module of the \textit{SciPy}\footnote{\href{https://www.scipy.org/about.html}{https://www.scipy.org/about.html}} software \citep{python_scypy}.
We subtracted this template from the LC, thus getting the residual LC (see the top two panels of Fig. \ref{fig:workflow}), which was the subject of the period search.

We ran tests on artificial data to examine whether the subtraction could introduce additional variation in the LCs. We generated EB LCs with \texttt{PHOEBE} 2.0 python code \citep{Prsa2016} using the periods and times of randomly chosen OGLE-IV EBs with additional noise.
Next, we carried out the same steps as earlier and applied the period determination method (see Sect. \ref{period_det}) to the residuals.  
The search concluded without finding any new significant periods and all detections could be traced back to the orbital period of the original EB. This implies that this method does not add artificial periodic variations to the data.

\subsection{Period determination} \label{period_det}
To determine the period of the residual LC we decided to use the phase dispersion minimisation \citep[PDM;][]{PDM} method. It finds periodic signals with non-sinusoidal wave-forms efficiently.
For this purpose we used the asynchronous PDM method of \texttt{cuvarbase}\footnote{\url{https://github.com/johnh2o2/cuvarbase}}, which allows the computationally intensive PDM
method to use the graphics card. This way we drastically reduced the run-time of our code.
After some initial tests, we chose the binned linear interpolation method of the PDM with the period range of $0.01-10$ days and with
20 phase bins, in frequency mode. For each period analysis we used 2 million test period values.

Furthermore, the period of every pulsating background candidate was also checked and validated by Lomb-Scargle (LS) periodograms \citep{VanderPlas_2018} using the \texttt{timeseries} module of \texttt{astropy} package \citep{astropy:2013,astropy:2018,astropy:2022}.

\subsection{Selection of variable star candidates}
When working with this high number of targets, it is worth searching for variable stars in an automated manner. For the automatic selection, we found that the following criteria were suitable in order to exclude non-variable stars: (1) the found period should not to be close to 1 day ($P < 0.98^d$ or $1.02^d<P$) to avoid the sampling period; (2) the period ratio of the candidate and the OGLE binary should not be close to 1; if it is, it is more likely to be an effect caused by stellar spots or the inappropriate removal of the EB signal; and (3) the maximum flux difference between the points of the FBLC has to be higher than the average of the standard deviations of the original data calculated for each phase bin.

In the case of Algol-type systems where the out-of-eclipse regions are relatively flat, we removed the eclipses from the LCs as an alternative check for background variations. First, we determined the eclipse borders with the method presented by \citet{Hajdu2022}, then we removed the points that belong to the eclipses. Finally, using the PDM and LS methods, we searched for periodic variations and compared the results with those obtained automatically.

\subsection{Tests with artificial LCs} \label{sec:test}
To test the reliability of our method, we generated an artificial dataset and performed the same procedure described above. 
We used the LCs of those EBs where we did not identify any background variability as the basis of the generation. We randomly selected $\sim30,\!000$ variable stars from the OGLE catalogues. To mimic the diversity of the real background stars, we used EBs, RR Lyraes, Cepheids, and $\delta$ Scutis and injected these LCs into the EB dataset. Using the periods from the OGLE variable catalogues, we phase-folded and binned the LCs of the variables. Then, we used the \texttt{UnivariateSpline} method (see Sect. \ref{sec:subtraction}) to fit the phase curves with a continuous model. Using the obtained phase-based curve, we determined the flux values at the time stamps of a randomly selected EB, and we added these values to the original LC together with the noise calculated from the residual phase-folded LC of the variable. In order to not let the background variable dominate the LC we selected variables where the amplitude is less than half of the amplitude of the `foreground' EBs.

\begin{figure}
    \centering
    \includegraphics[width=\linewidth]{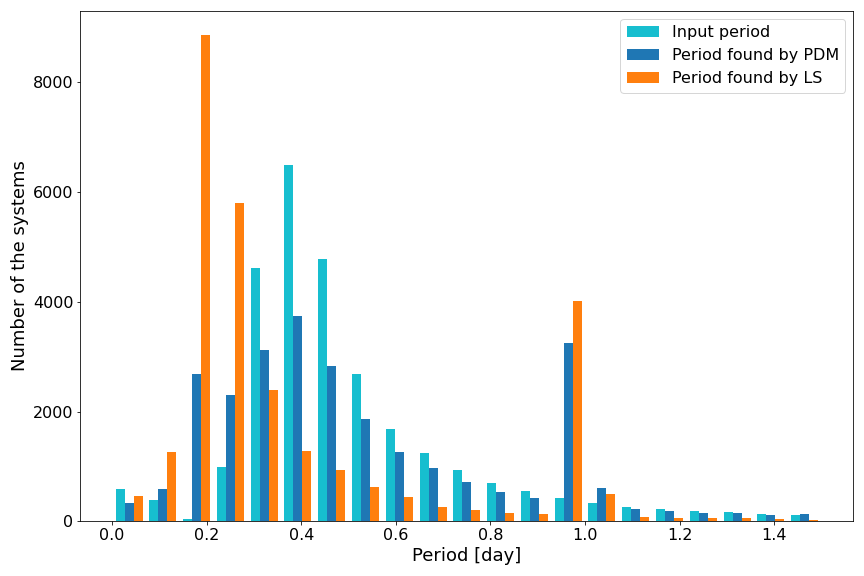}
    \caption{Distribution of the injected periods (cyan) of background variables used to validate the automatic method and the periods found by PDM (blue) and LS (orange) algorithms.}
    \label{fig:test_per_dist}
\end{figure}

The results of our validation test are presented in Fig.~\ref{fig:test_per_dist}, where we show the distribution of the input periods and the recovered periods produced by both period search methods.
The distribution of the PDM periods is similar to the input periods, except the additional peak at around 1-day, which is caused by the daily aliasing in ground-based observations. The peak of the distribution of the LS periods is shifted towards lower values. LS periods are concentrated around the half of the input periods, which is caused by the injected EB systems, where this method found the half of the orbital periods. The conspicuous peak at around 1 day is caused by the sampling effect, similar to the case of the PDM analysis.

\begin{figure}
    \centering
    \includegraphics[width=\linewidth]{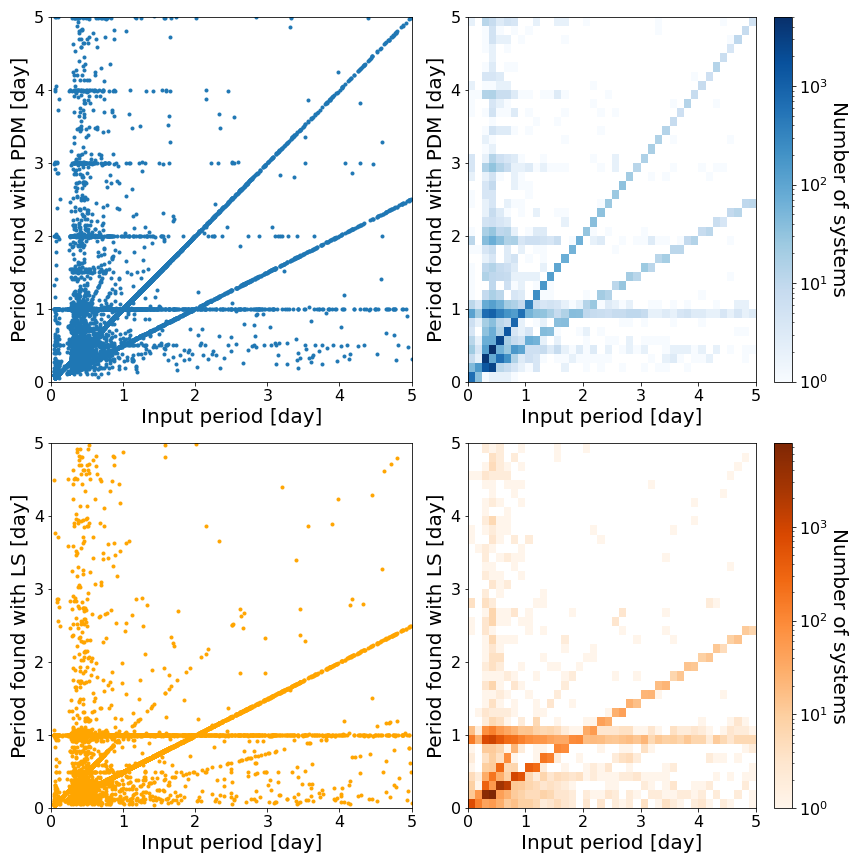}
    \caption{Comparison of the injected and recovered periods for PDM (top row; shades of blue) and LS (bottom row; shades of orange) analysis. The results are plotted on a $P_{\textup{input}} - P_{\textup{found}}$ plane. To highlight all aspects, the left panels show all individual points, while the right panels present logarithmic number densities.}
    \label{fig:per_vs_per}
\end{figure}

To further compare the two methods we plotted the input and found periods against each other in Fig. \ref{fig:per_vs_per}.
The left panels show all individual data pairs (input period and identified period per LC). The right panels show the same but after binning the data and then shaded with the logarithmic number density of the points. This way we can highlight areas with high densities in the plots that are not apparent on the left side. 
These figures, like Fig.~\ref{fig:test_per_dist}, clearly show that the PDM method finds the correct variability period in the majority of cases, whereas the LS very frequently identifies half (or one quarter) of the orbital periodicity as the dominant cycle length for EBs.
The multiplies of the daily aliasing can easily be spotted on the left images as horizontal lines, though it is only significant for the 1-day period if we consider the number densities.

\begin{figure}[ht!]
    \centering
    \includegraphics[width=0.9\columnwidth]{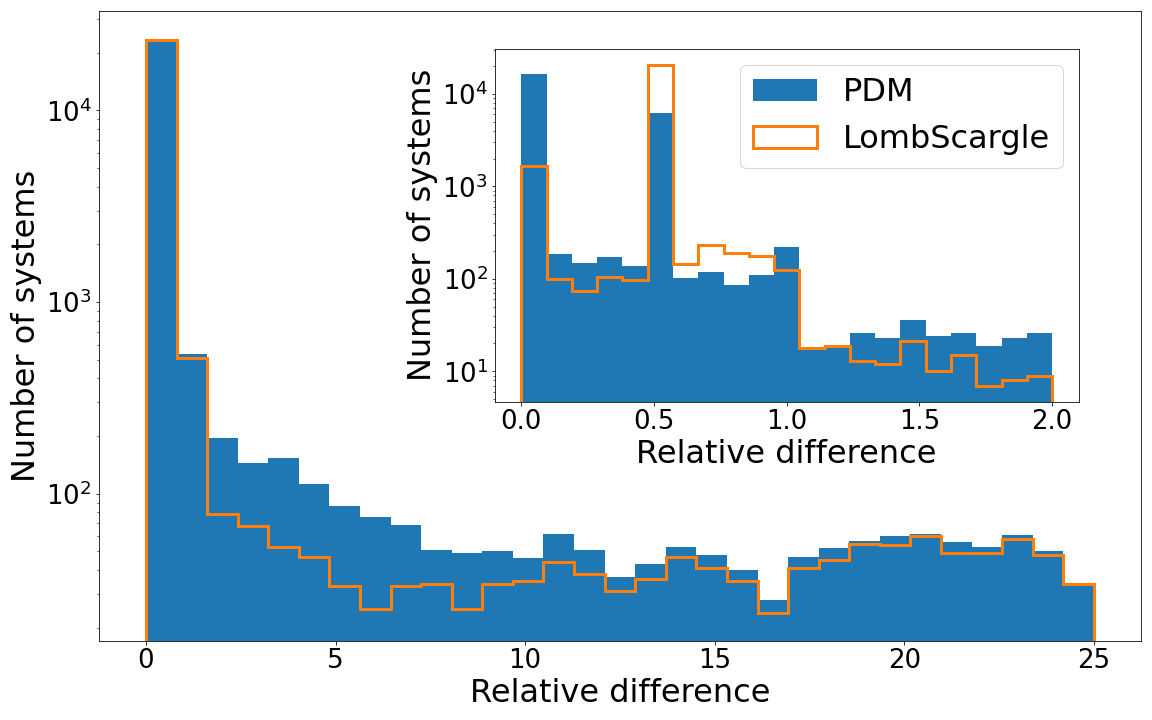}
    \caption{Distribution of the relative differences $\Big(
    \left|P_{\textup{input}}-P_{\textup{found}}\right|/ P_{\textup{input}} \Big)$ between the input and found periods for the PDM (blue) and LS (orange) methods. Note that the vertical scale is logarithmic. To highlight the precision of the methods, the insert shows the $[0.0,2.0]$ region. Here, the peaks around zero and 0.5 are two orders of magnitude higher than any of the other bins.}
    \label{fig:per_dist}
\end{figure}

In Fig. \ref{fig:per_dist} we compared the precision of the two period detection methods, where we plotted the distribution of the relative differences
$\Big(\left|P_{\textup{input}}-P_{\textup{found}}\right|/ P_{\text {input}} \Big)$ between the input and found periods in a logarithmic scale. It shows significant peaks around zero and 0.5, which are two magnitudes higher than the other bins in the insert image where the relative difference ranges from $0.0$ to $2.0$. Again, the peak at 0.5 bin comes from detecting half of the orbital periods for EBs.

After running our period search algorithms on the artificial dataset, we can summarise the followings. The cases where the found period is equal to the input period or the half or the double of it (within 0.1\%) is more than 75\%.
15\% of the failed outputs comes from variables where the period search was misled by the observational sampling (1-day and its multiples), which can be easily filtered out by appropriate criteria. The last 10\% is a result of failed disentangling and quasi-periodic noises.

Consequently, we conclude that the periods of the background signals can be recovered in most cases and the reliability of our algorithms is confirmed by the previous plots. Furthermore, we can state that the PDM method is more suitable for the identification of periodic variables, and it has already been proven very effective at finding EBs \citep{2014Mayangsari,LaCourse2015,2021Bienias,Botan2021}, which, based on the distribution of different variable types in the original OGLE data, are expected to form a significant part of the background candidates. Nonetheless, all PDM periods were also confirmed by the LS analysis.

\subsection{Eclipse timing and LTTE modelling} \label{method_lte}

Common variations in the $O-C$ diagrams of the eclipses and/or pulsations would indicate that the variations belong to the same stellar system.
We used the method presented by \cite{Hajdu2022} to automatically create the $O-C$ diagrams from the LCs of the investigated systems and their background components after disentangling them the same way as presented in Sect. \ref{sec:subtraction}.
After creating the $O-C$ diagrams and identifying any multiple systems, we aimed to fit the light-travel-time effect (LTTE; e.g. \citealt{Irwin1952}) to determine the values of the common orbital parameters by using Eq. (2) of \cite{Borkovits_2015}:
\begin{equation}
    \Delta_{\textup{LTTE}}= -\frac{a_{\textup{AB}}\cdot \sin i_2}{c} \frac{(1-e_2^2) \sin (v_2 +\omega_2)}{1+e_2 \cos v_2},
    \label{eq:o-c}
\end{equation}
where $\Delta_{\textup{LTTE}}$ corresponds to the $O-C$ values. 
In Eq. (\ref{eq:o-c}) $a_{\textup{AB}}$ is the semi-major axis of the close binary around the centre of mass of the triple system, while $e_2$, $\omega_2$, $i_2$ and $v_2$ stand for the eccentricity, the argument of periastron,  and the inclination of the relative outer orbit and the true anomaly of the third component, respectively. We note that for these systems, whose period ratio is relatively high ($P_2/P_{\textup{EB}} \gtrsim 100^d$), the dynamical perturbations are negligible (\citealt{Borkovits2022PinPout}).

To fit Eq. (\ref{eq:o-c}) to the derived $O-C$ values and hence calculate the outer orbital parameters, we used the python \texttt{emcee}\footnote{\url{https://emcee.readthedocs.io/en/stable/}} package, which is an implementation of the Markov-chain Monte Carlo  method \citep{F-M_Hogg_2013}.
For each multiple stellar system candidate, we also calculated the mass function of the tertiary component based on the LTTE parameters with the well-known method as
\begin{equation}
    f(m)=\frac{4 \pi^2 a^3_\mathrm{EB} \sin^3i_2}{G P_2^2},
    \label{eq:mass_f}
\end{equation}
where $a_{\textup{EB}}\sin i_2$ is the projected semi-major axis and $P_2$ is the period of the LTTE orbit, and $G$ is the gravitational constant.

\section{Classification} \label{classification}
\subsection{Light curve analysis}

\begin{figure*}
    \centering
    \includegraphics[width=\linewidth]{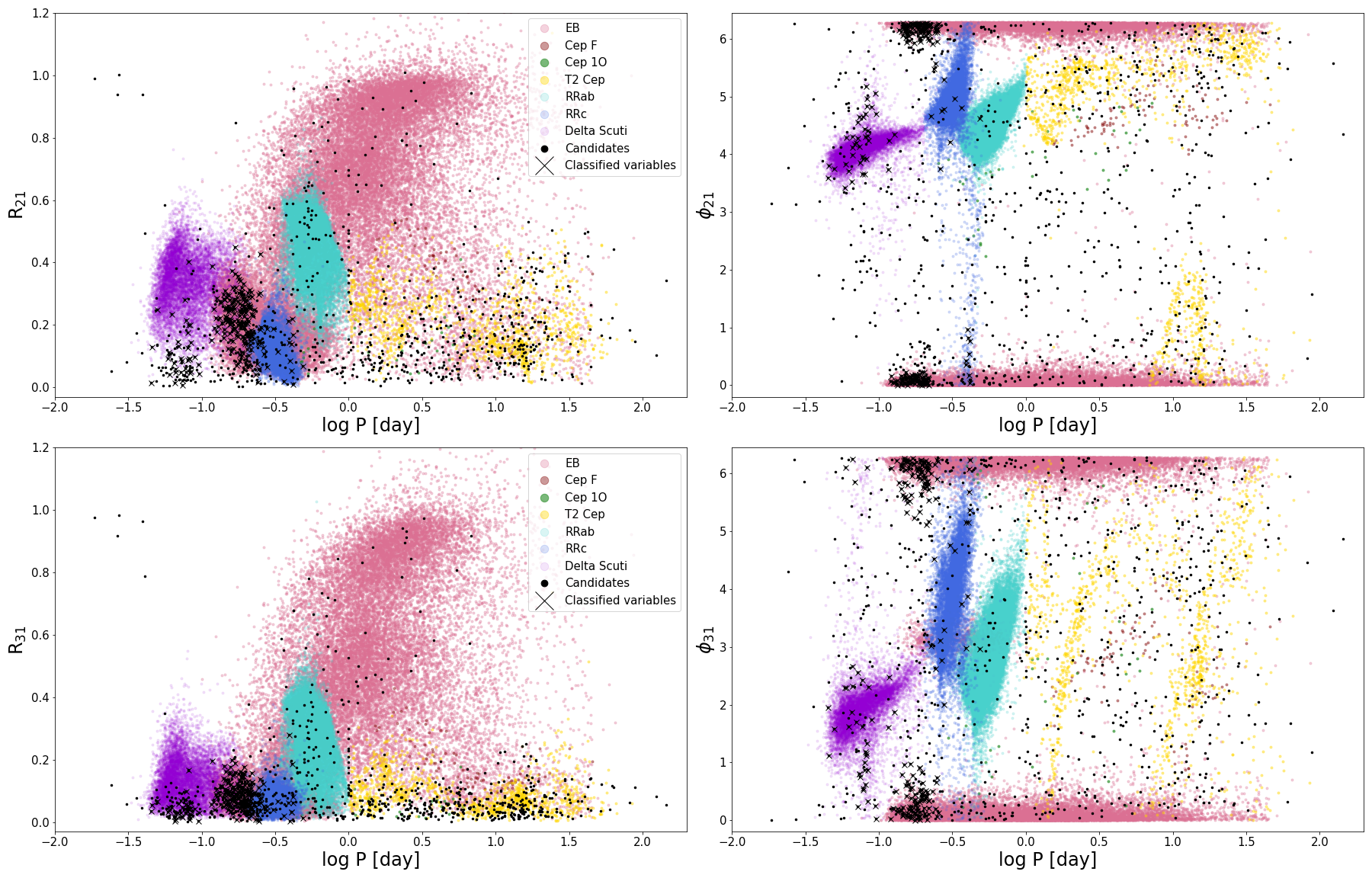}
    \caption{Relative Fourier parameters of the variable candidates we found (black dots) and classified variables (black Xs).
    The colourful background is composed of the known OGLE variable stars: EBs (light pink), fundamental-mode Cepheids (light red), first-overtone Cepheids (light green), type II Cepheids (yellow), RRab (turquoise), RRc (royal blue), and $\delta$~Scuti stars (dark violet).}
    \label{fig:fourier_all}
\end{figure*}

First, we visually inspected the residual LCs, and selected more than 90 EBs according to the phenomenological properties of their LCs.
To classify the rest of the candidates we used three methods: we checked the results manually, calculated the relative Fourier parameters of the LCs \citep{Simon_Lee_1981}, and tested an image-based machine learning classification method \citep{2020Szklenar,Szklenar22}. For the last method we used the phase-folded LCs and the periods. The predicted variable types with the highest scores are listed in Table \ref{table:class_image}.
We assigned the most probable variability type when the score exceeded 80\%, and the two most likely types otherwise.

We computed the first four relative Fourier parameters $R_{i1}=A_i/A_1$ and $\phi_{i1}=\phi_i-i\phi_1$ with the Fourier parameters of the frequency of the main variation and its harmonics fitted. Since the relative Fourier parameters that are available in the OGLE database were calculated using cosine-based Fourier series, we also used series in the form of 
\begin{equation}
    m(t) = m_0 + \sum_i A_i \cos(2\pi i f t + \phi_i).
    \label{eq:fourier}
\end{equation}
In Eq. (\ref{eq:fourier}) $m_0$ is the average brightness, $f$ is the dominant frequency, $i$ is the order of the peak in the harmonic series, $A_i$ and $\phi_i$ are the amplitudes and phases of the given frequency component, respectively. These quantities characterise the shapes of LCs. 

After calculating the relative Fourier parameters, we plotted them over the values of the known OGLE variables in Fig. \ref{fig:fourier_all} \citep{OGLE_rrl_2014,OGLE_cep_2017,OGLE_BLG,OGLE_dsct_2020}. We classified solely those candidates that fitted to exactly one variable type regarding all four relative Fourier parameters. Since the short period ($P\lesssim1^d$) section was clearer, in this work we focus on that segment of the plot.

\subsection{Contamination by known variables} \label{sec:contamination}

Since the Bulge region is a crowded area on the sky, where the blending phenomenon is a known systematic effect \citep{2021Melchior}, it is worth comparing our variable candidates to the already known and published variables (EBs: \citealt{OGLE_BLG}, $\delta$ Scutis: \citealt{OGLE_dsct_2020}, RR Lyraes: \citealt{OGLE_rrl_2014} and Cepheids: \citealt{OGLE_cep_2017}).
To verify that the found signal is not only contamination caused by a known variable or system, we compared the background periods found in the systems with the periods of the variables within their vicinity, to a maximum distance of 100".

We identified those systems as contaminated ones where the period difference was less than 0.05 days from a known variable within this distance range.
These systems are listed in Table \ref{table:contaminated} with the period of the foreground EBs and the background variables.

Taking the contamination into account, the plots used for classification are presented in Appendix \ref{sec:app_class_figures}. We made different figures for each variability type, colouring only a single type of the known variables. In the next step, we selected the group of stars that fitted into one type regarding the $R_{21}$ parameter, then excluded those that did not fit the other three parameters. Finally, we checked whether we found a known variable: the results of contamination by catalogued variables are marked with crosses in these figures.

\renewcommand{\arraystretch}{1.2}

\begin{table}
\caption{EB systems whose photometry contains a signal of a known background variable (BV). The full list can be found in Table \ref{table:cont_full}.}
    \centering
    \begin{tabular}{c l}
OGLE ID of EB & \multicolumn{1}{c}{OGLE ID of BV} \\\hline
OGLE-BLG-ECL-043403 & OGLE-BLG-RRLYR-01193 \\
OGLE-BLG-ECL-050415 & OGLE-BLG-DSCT-01176 \\
OGLE-BLG-ECL-140255 & OGLE-BLG-RRLYR-04853 \\
OGLE-BLG-ECL-162653 & OGLE-BLG-RRLYR-31535 \\
OGLE-BLG-ECL-164054 & OGLE-BLG-DSCT-03914 \\
OGLE-BLG-ECL-166633 & OGLE-BLG-RRLYR-31656 \\
OGLE-BLG-ECL-168002 & OGLE-BLG-ECL-168013 \\
OGLE-BLG-ECL-168013 & OGLE-BLG-ECL-168002 \\ 
... & \multicolumn{1}{c}{...} \\ \hline
    \end{tabular}
    \label{table:contaminated}
\end{table}
   
\section{Results} \label{result}

As a result of our search, we found 1190 systems whose residual LCs showed periodic variations. After the first visual inspection of the LCs, we identified more than 90 EBs. We classified the rest using the relative Fourier parameters and decided to focus on the short period section; this left us with 354 candidates.

\begin{figure}
    \centering
    \includegraphics[width=\linewidth]{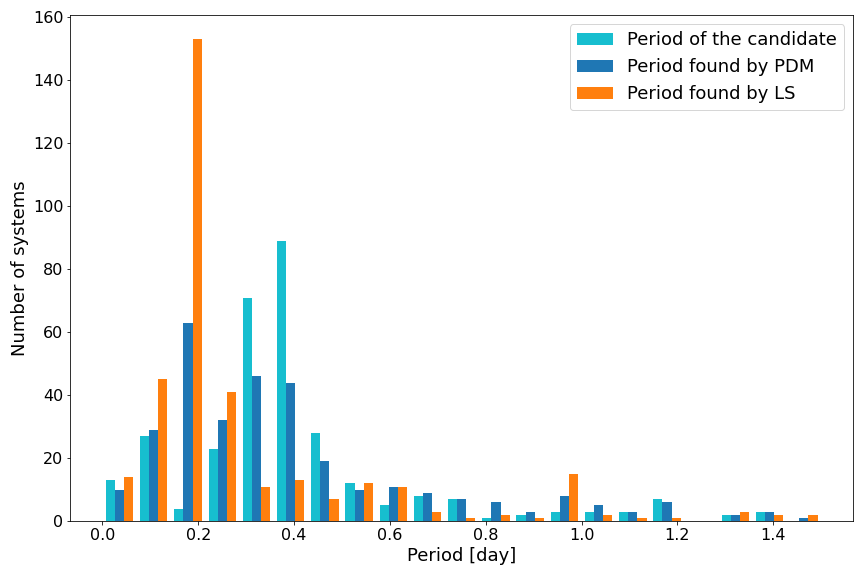}
    \caption{Distribution of the periods of the candidates (cyan) and the periods found by PDM (blue) and LS (orange) algorithms.}
    \label{fig:candidates_per_pdm_ls}
\end{figure}

After visually inspecting the LCs and the fine-tuning the period values of the candidates, we compared the distribution of their periods with those found by PDM and LS methods (as in Sect. \ref{sec:test}). The PDM method is in better agreement here as well, as seen in Fig.~\ref{fig:candidates_per_pdm_ls}. The distribution of the LS periods shows that the LS method usually finds the half period in the case of the EBs. This phenomenon can be identified for the PDM method too, but at a much smaller scale.

Lastly, we excluded the ones that were results of contamination by known OGLE variables from our catalogue.
This left us with 292 new variable stars, which are listed in Table \ref{table:var_full}.

\begin{figure}
    \centering
    \includegraphics[width=0.9\columnwidth]{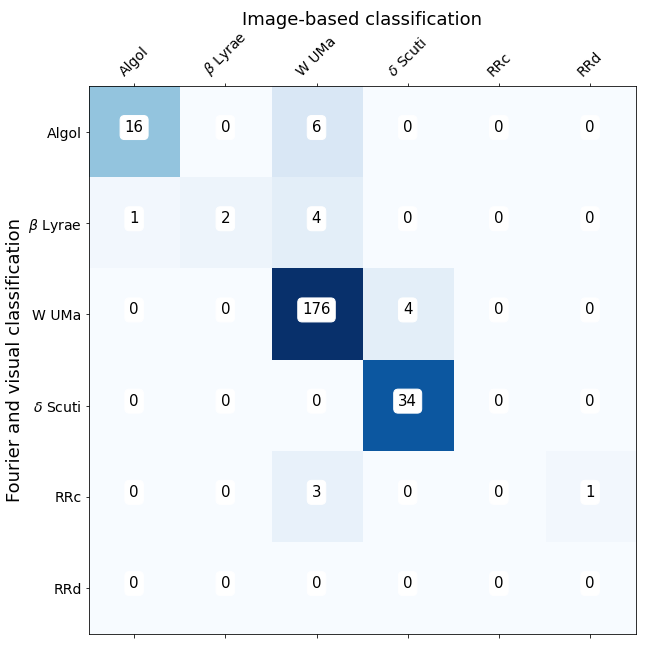}
    \caption{Confusion matrix of our new catalogue and the results of image-based classification where the highest probability exceeded 80\%.}
    \label{fig:conf_matrix}
\end{figure}

As we mentioned earlier, we also used an image-based machine learning classifier, whose results are in close agreement with the Fourier-parameter-based classifications. We cross-matched the results of the different classification methods for the objects where the image-based machine learning method resulted in a classification score above 80\% (247 stars). The comparison is presented in Fig.~\ref{fig:conf_matrix}.
We note that there is a 91.89\% match in the case of $\delta$ Scuti variables, and the numerous W UMa type has the second best rate with a 91.19\% match.
The examined subset of classified variables is in near-perfect agreement with our catalogue (92\% match). 
Though, as Fig.~\ref{fig:conf_matrix} shows, classifying $\beta$ Lyrae stars according to their LC shape might be the biggest challenge, since it is not an easy task to determine whether the primary and secondary minima have the same depth or where the borderlines of an eclipse are.
The classification of RRc stars also seems to cause great difficulty. We identified 13 variables as RRc stars, while the image-based classifier flagged only four of them with a high probability but not as the same type in any case. These four stars are in the outer regions of RRc stars considering their position in Fig. \ref{fig:fourier_rrc} and therefore their classification with Fourier parameters might be ambiguous, meanwhile the classifier could also have reached its limits.

\begin{figure}
    \centering
    \includegraphics[trim={3.5cm 1cm 3.5cm 1cm}, clip, scale = 0.5]{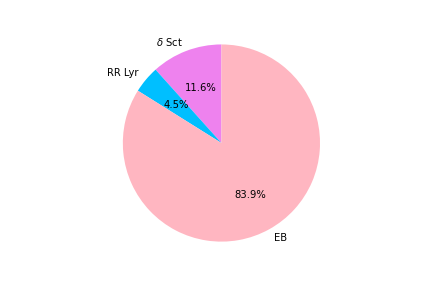}
    \includegraphics[trim={3.5cm 1cm 3.5cm 1cm}, clip, scale = 0.5]{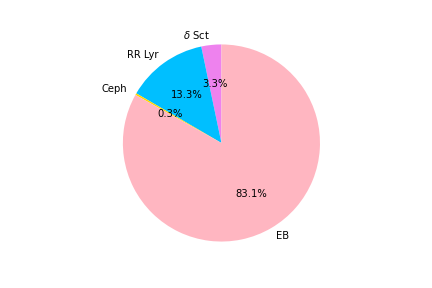}
    \caption{Distribution of variables found through our study (left) and by the OGLE team (right) towards the Bulge. The relative abundance of EBs is quite similar. The visible difference between the pulsating variables is due to their luminosities, since $\delta$ Scuti stars are comparable with EBs in terms of brightness.}
    \label{fig:dist}
\end{figure}

The distribution of these variables is shown in Fig. \ref{fig:dist}, along with a comparison of the relative abundances of known OGLE variables towards the Bulge as well.
The most notable difference between the diagrams is in the rates of RR~Lyrae and $\delta$ Scuti stars. The OGLE team identified 8.8\% more RR~Lyraes, while we found 8.3\% more $\delta$ Scuti pulsators. This phenomenon is a result of the characteristics of the variable types. As RR~Lyrae stars are on the horizontal branch, they are more luminous and would outshine the EB they reside in, and hence the OGLE team could have found them. On the other hand, $\delta$ Scuti stars are main sequence or post-main sequence stars and therefore their brightness is comparable with EBs. In fact the selection function is dominated by the amplitude measured in flux generally, which, in the case of $\delta$ Scutis, is comparable with EBs.

\begin{figure}
    \centering
    \includegraphics[width=\columnwidth]{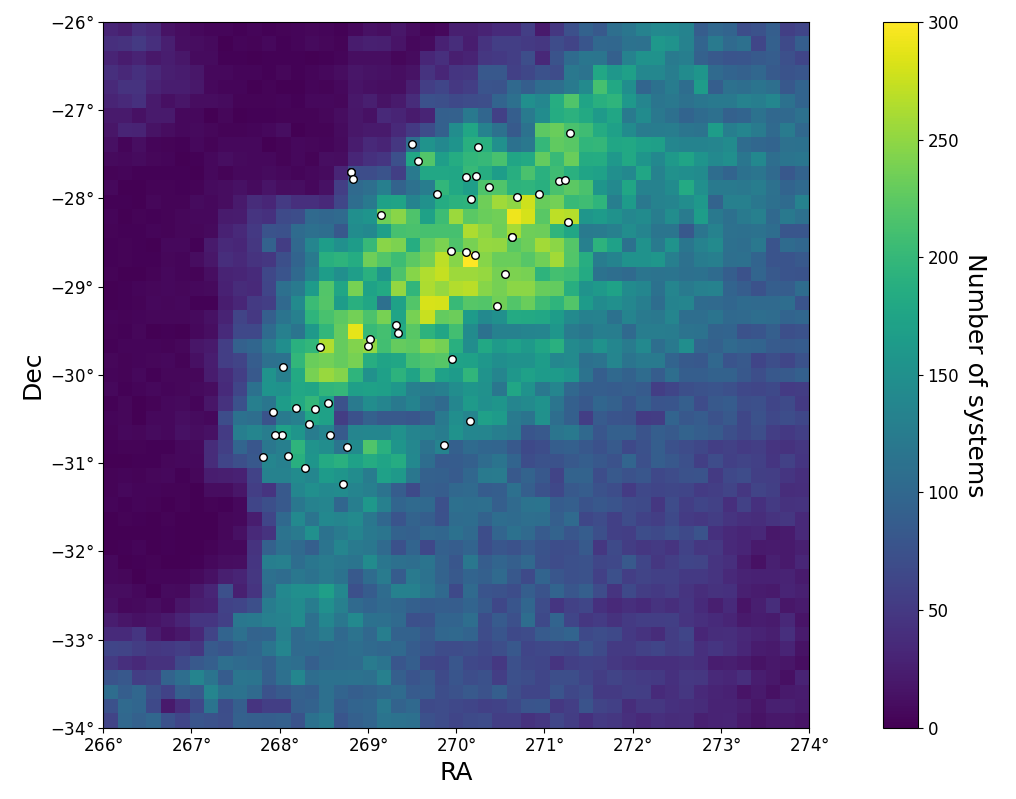}
    \caption{Newly found RRc and $\delta$ Scuti stars (large dots) plotted over the distribution of OGLE EBs. The colour scale shows the number of EBs per bin. One bin corresponds to an area of $10'\times10'$.}
    \label{fig:rrc_dsct_pos}
\end{figure}

We tested where the newly found short-period pulsating variables are compared to the Bulge. We analysed their positions against the EB catalogue through Fig. \ref{fig:rrc_dsct_pos}.
If we take into account that the region closest to the disk is highly affected by dust, and therefore neglect this region, we can state that for our search to be successful we needed at least one EB per square arcminutes. But we identified a significant number of new variables where the mean density of EBs was around 1.48 arcmin$^{-2}$ or higher. According to the \textit{Gaia} Data Release 3 EB catalogue \citep{gaia-EB-2022}, the Bulge has the highest density of EB stars, exceeding 2000 deg$^{-2}$ or 0.55 arcmin$^{-2}$, and therefore we do not expect a similar frequency of blended variables elsewhere in the Milky Way.

\subsection{\texorpdfstring{$\delta$}{d} Scuti variables}
\begin{figure}[ht]
    \centering
    \includegraphics[width=0.95\linewidth]{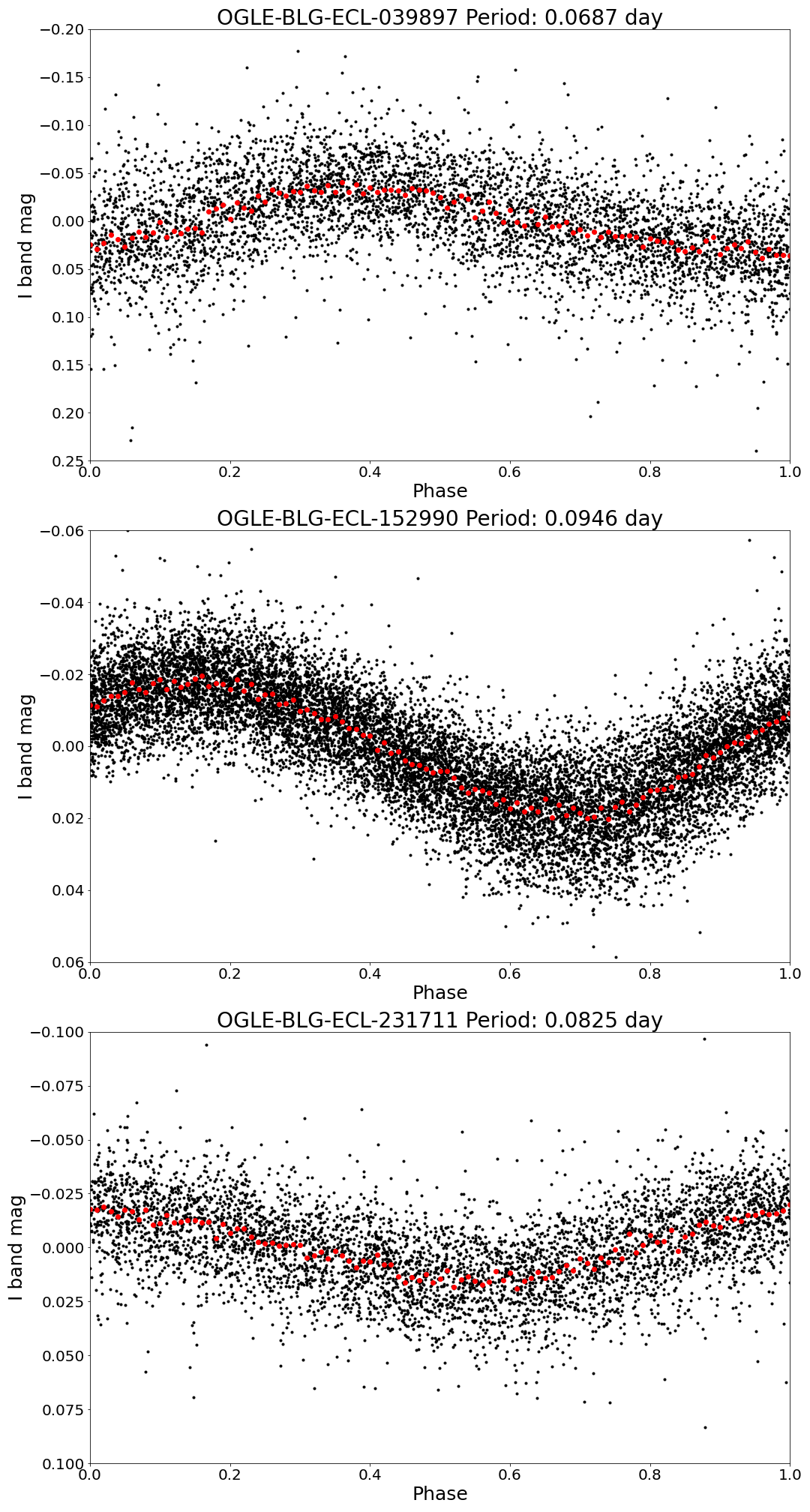}
    \caption{LCs of three $\delta$ Scuti stars from our catalogue. The red dots mark the FBLC.}
    \label{fig:dsct_lc}
\end{figure}

We identified 34 new $\delta$ Scuti variables. As shown in Fig.~\ref{fig:fourier_all}, all of them have low R$_{21}$ and R$_{31}$ values compared to their already catalogued counterparts. These LCs are similar to the LCs of EBs, which aligns with our expectations, since otherwise the OGLE team would have been able to identify them. Three example LCs are presented in Fig.~\ref{fig:dsct_lc}.

\subsection{RR Lyrae variables}
\begin{figure}[ht]
    \centering
    \includegraphics[width=0.95\linewidth]{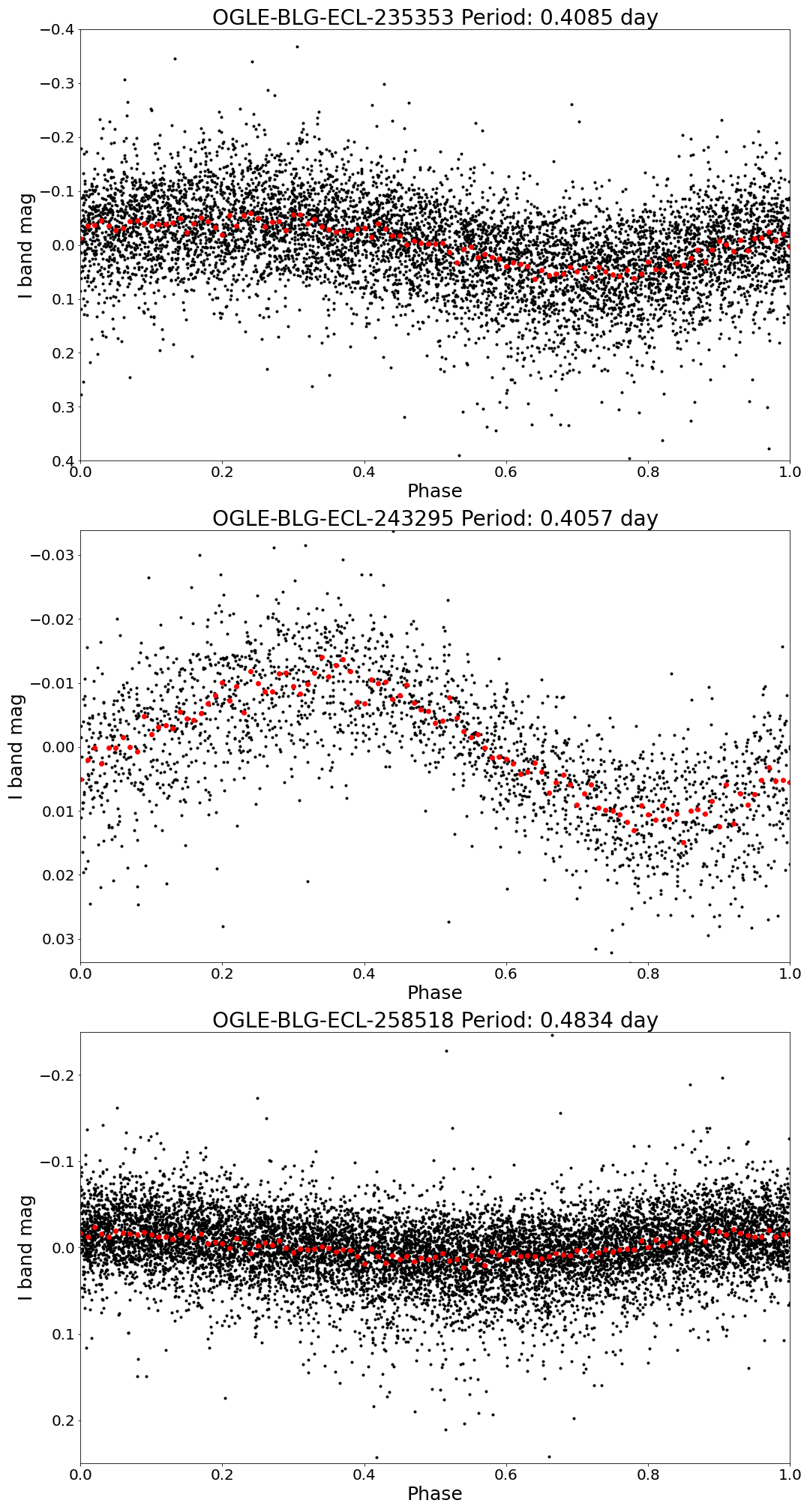}
    \caption{LCs of three RRc stars from our catalogue. The red dots denote the FBLC.}
    \label{fig:rrc_lc}
\end{figure}

We discovered 13 new RR Lyrae stars. As we mentioned before, a low number of this type is reasonable since they are luminous and not anticipated in close binary systems; therefore, we only expect chance alignments here.
The classification of these variables was also made according to their relative Fourier parameters (see Fig. \ref{fig:fourier_rrc}). As it turned out, all of them are RRc variables, first overtone pulsators with shorter periods, as all of our RRab candidates (ten systems) have already been identified by the OGLE survey.
We present a few LCs of found variables in Fig.~\ref{fig:rrc_lc}.

OGLE-BLG-ECL-258518 deserves special recognition as it remains a puzzling case even after the manual check of its LC and period. It might be a very long period RRc (as we have classified it) or an RRab star with a considerably flat LC, as can be seen in the bottom panel of Fig.~\ref{fig:rrc_lc}. 

\subsection{Eclipsing binaries}

Eclipsing binaries take up a significant part of variable stars; therefore, it is not surprising that we found hundreds (246) of them in the backgrounds of the examined systems. We analysed the period and the morphology parameters ($C$) \citep{2012Matijevic} of the systems. The latter were determined by the python package \textit{polyfits} used for calculating the $C$ parameters for the OGLE binaries \citep{Bodi2021}. 

\begin{figure}
    \centering 
    \includegraphics[width=0.9\columnwidth]{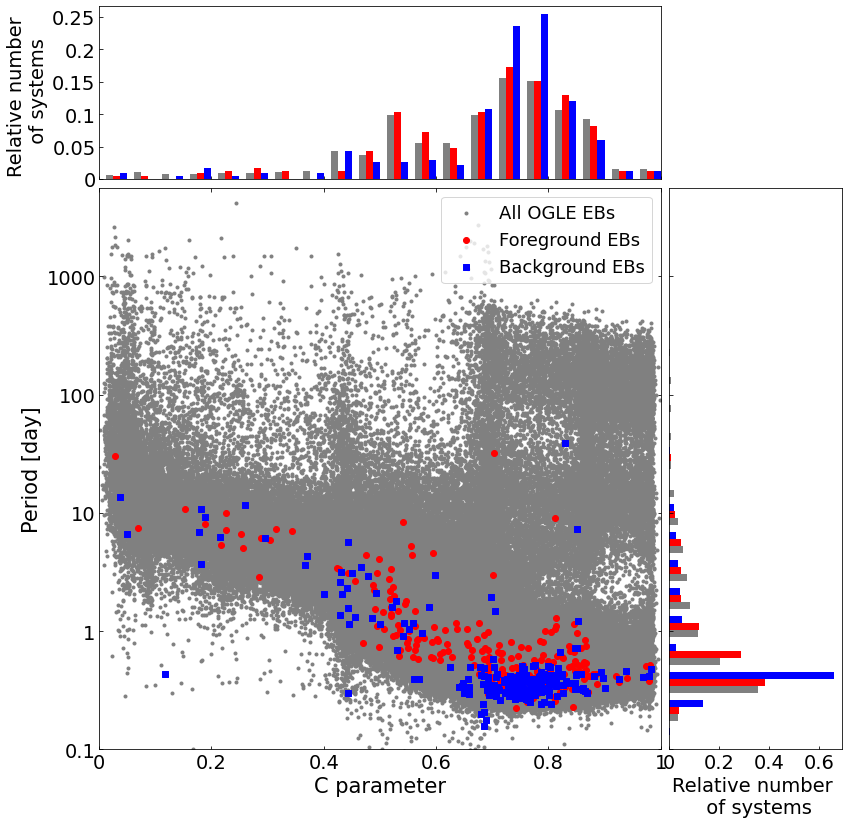}
    \caption{Period-morphology distribution of the OGLE EBs, the foreground (red) and the background (blue) EBs. The foreground binaries mimic the distribution of OGLE binaries surprisingly well, and the distribution of the found background systems shows a stronger peak in a well-defined phase range ($0.6 < C < 0.9 $ and $0.2 < P < 0.5$ ).}
    \label{fig:period-morphology}
\end{figure}

The period-morphology distribution of the systems is presented in Fig.~\ref{fig:period-morphology}.
Since most identifications come from the Fourier amplitudes where we only looked at the short-period range, the background EBs mainly group there ($P<1^d$). Consequently, the period distribution has a more significant peak compared to the distribution of the total OGLE list.
In general, the distribution of the $C$ parameter follows the trend of the OGLE list, although the bump in the range of $0.4<C<0.6$ is notably flatter.

As we focused on the short period section of Fig.~\ref{fig:fourier_all}, it is no surprise that largest part of EBs that we found consists of W~UMa-type stars (194). We also identified 11 new $\beta$ Lyrae and 41 Algol-type variables.

\subsection{Quadruple systems from eclipse timings}
In addition to the basic parameters, it is worth examining whether there is any physical connection between the systems, as it was done by \cite{Zasche2019}, who analysed 146 doubly eclipsing systems. Following their footsteps, we also generated the $O-C$ diagrams of the systems, for which we used the cleaned LCs. Our method is detailed thoroughly in Sect. \ref{method_lte}.

\begin{figure*}
    \centering
    \includegraphics[width=\columnwidth]{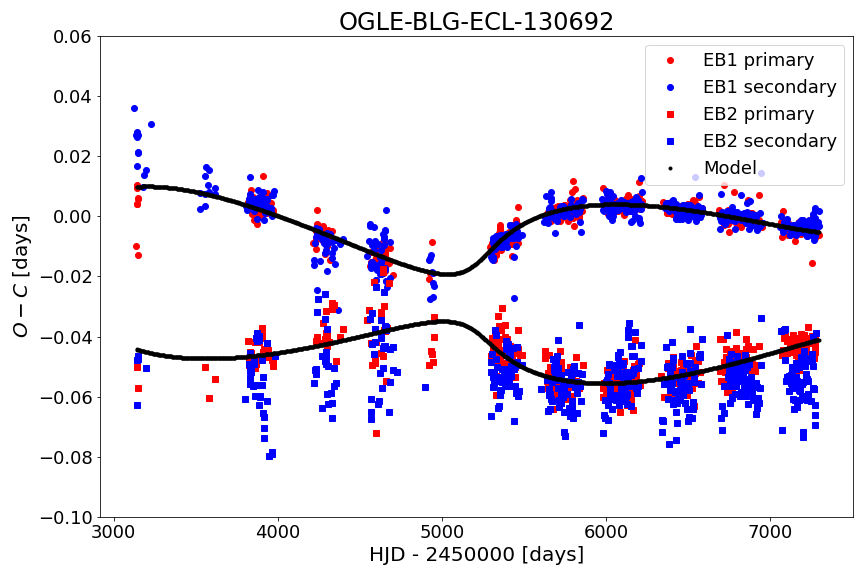}
    \includegraphics[width=\columnwidth]{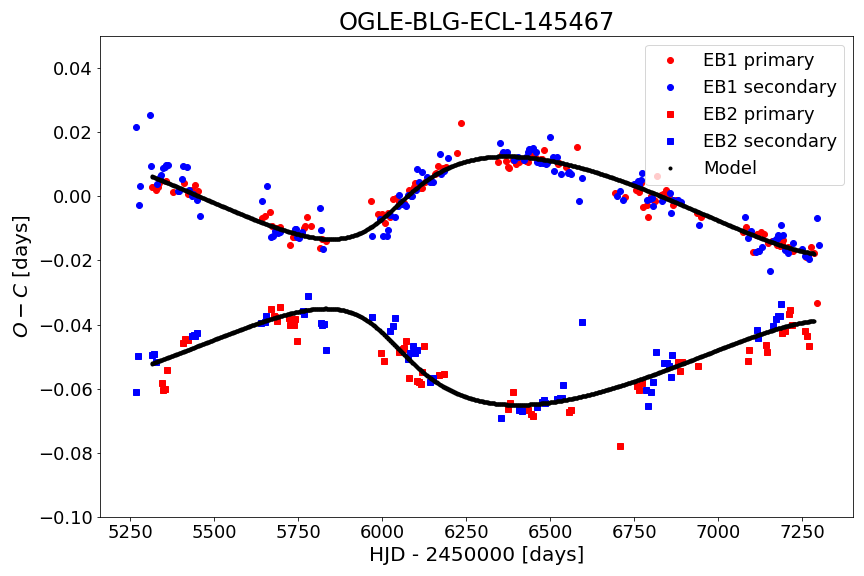}

    \includegraphics[width=\columnwidth]{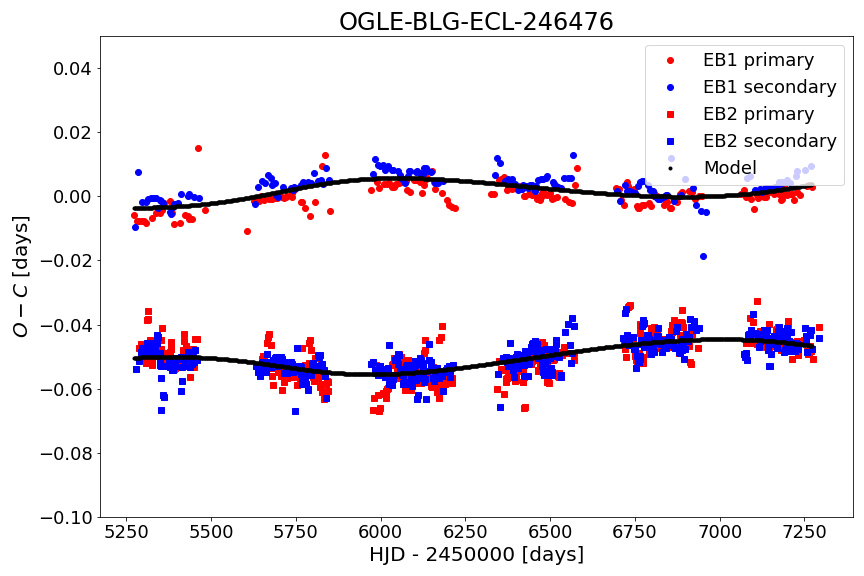}
    \includegraphics[width=\columnwidth]{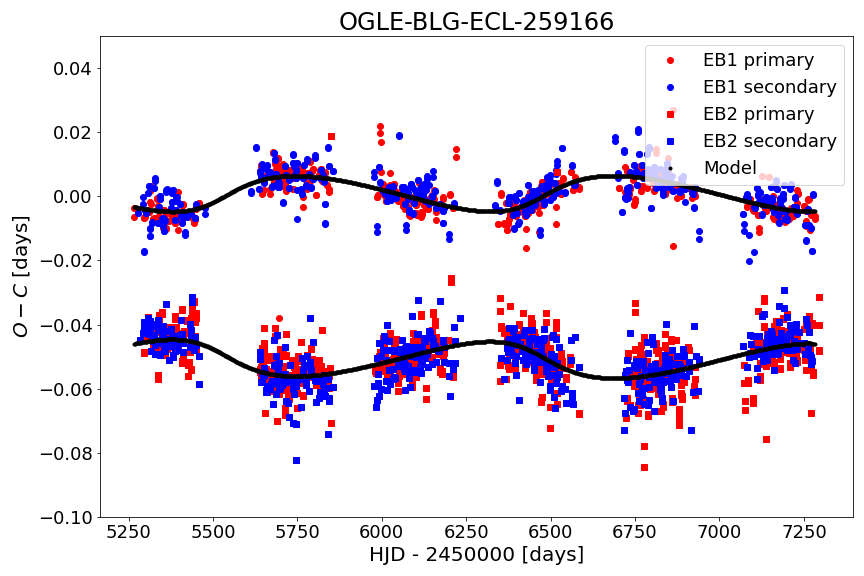}
    
    \includegraphics[width=\columnwidth]{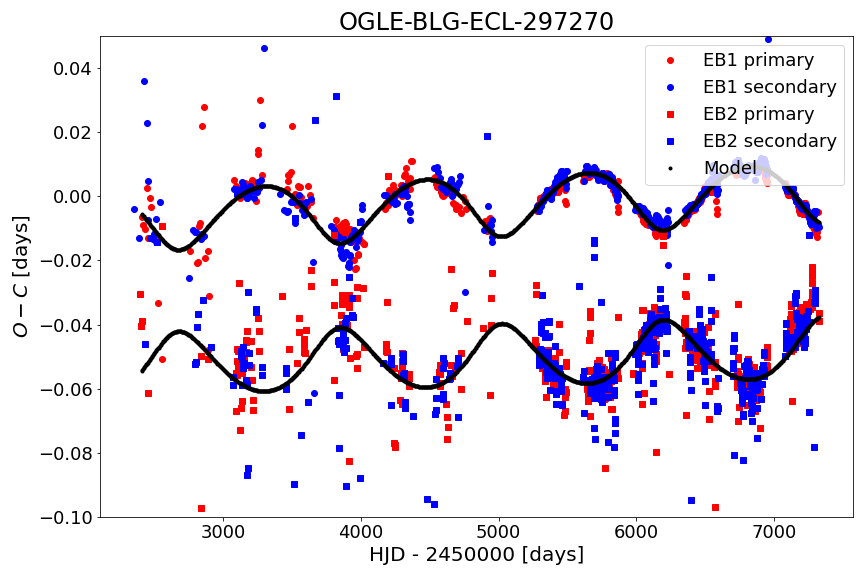}
    \includegraphics[width=\columnwidth]{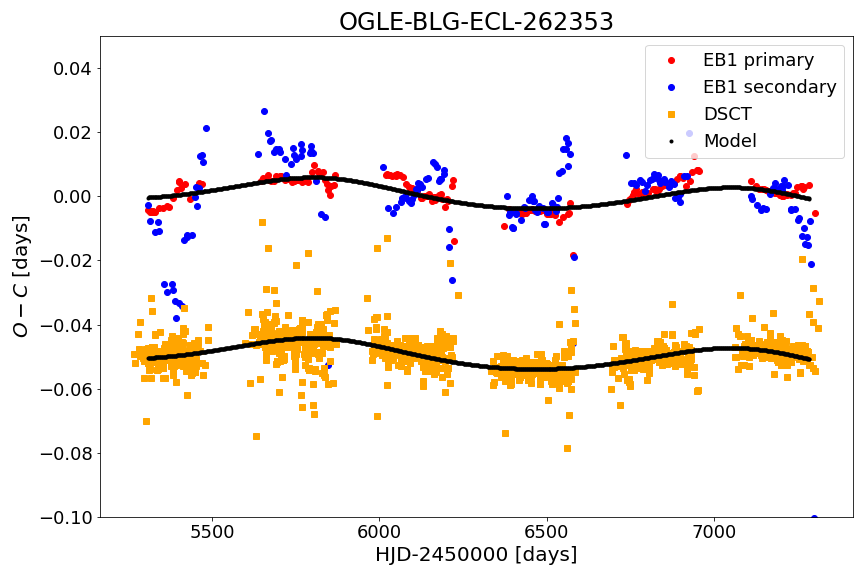}

    \caption{ETVs of our candidate double EB systems and the triple candidate system with a $\delta$ Scuti pulsator (bottom right). For better visual transparency, we shifted the $O-C$s of the background EBs and the $\delta$ Scuti by $0.05$ days. The dots represent the $O-C$ diagram of the foreground binary, while the squares represent the $O-C$ diagram of the background binary. The red and blue colouring indicates the ETVs of the primary and secondary minima, respectively. The fitted LTTE is shown as a black curve. In the case of the triple star candidate system, the timing variations of the $\delta$ Scuti is marked with orange dots.}
    \label{fig:all_double_EB_oc}
\end{figure*}

\begin{figure*}
    \centering
    \includegraphics[width=\columnwidth]{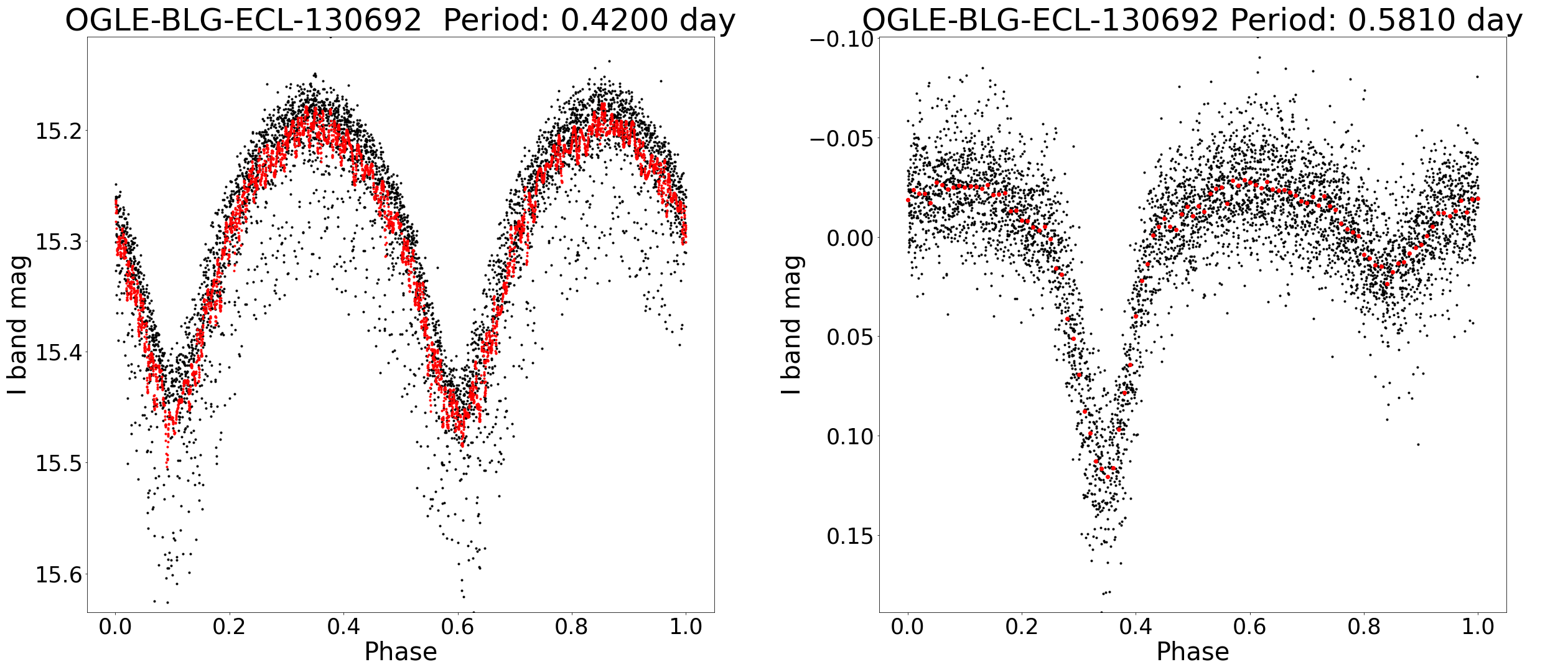}
    \includegraphics[width=\columnwidth]{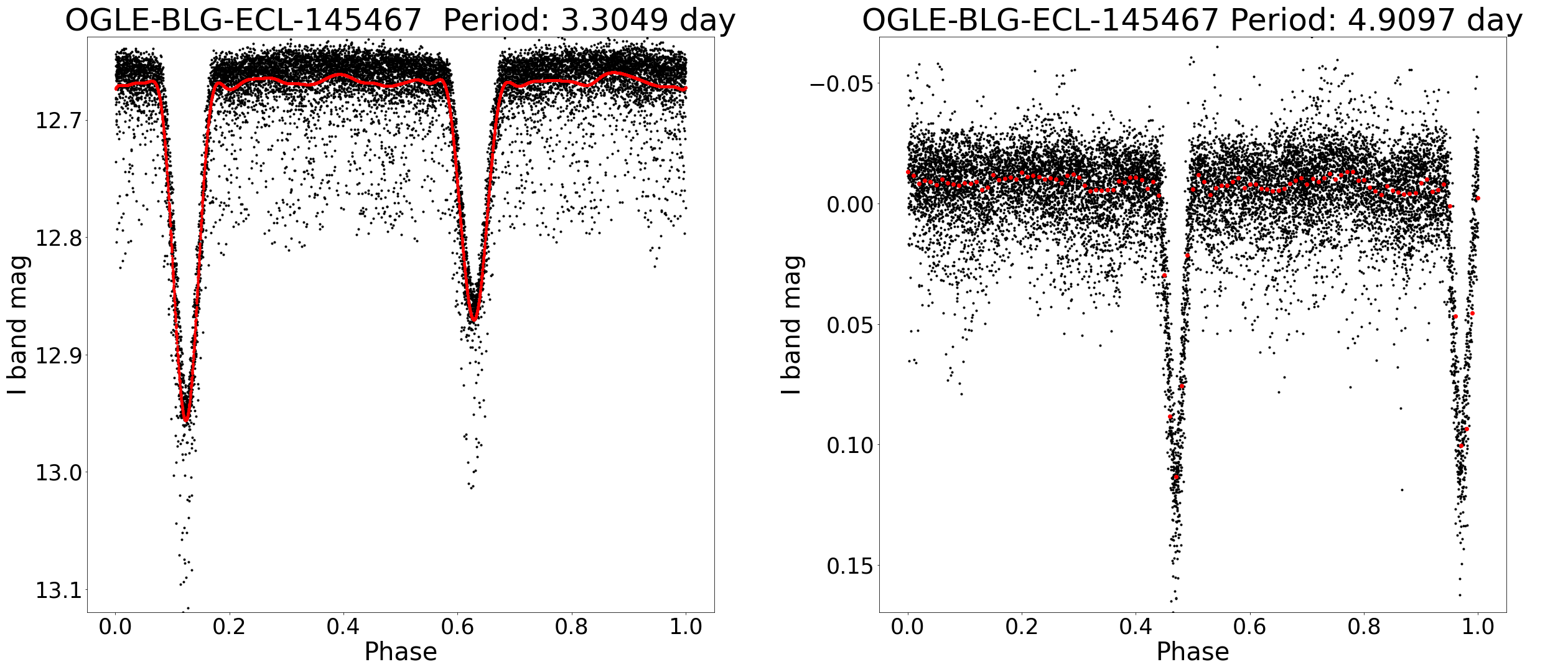}
    \includegraphics[width=\columnwidth]{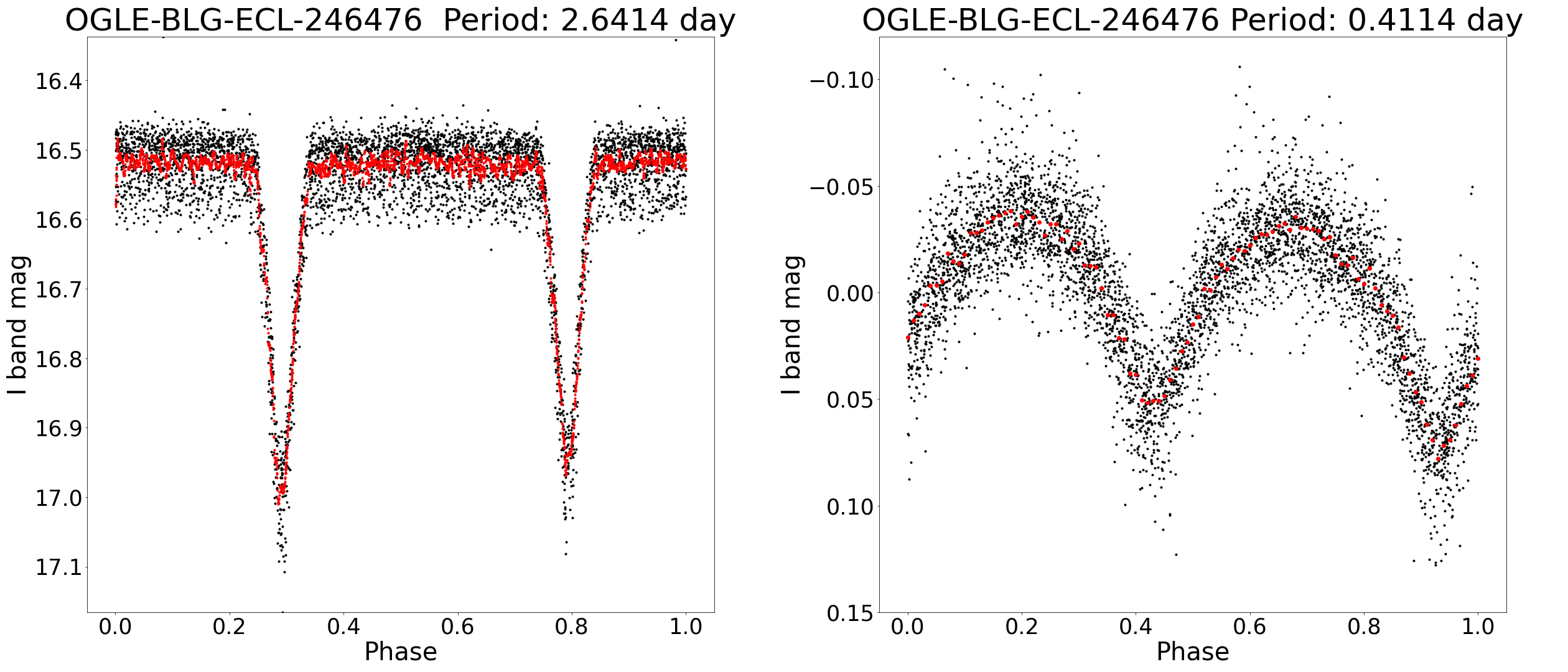}
    \includegraphics[width=\columnwidth]{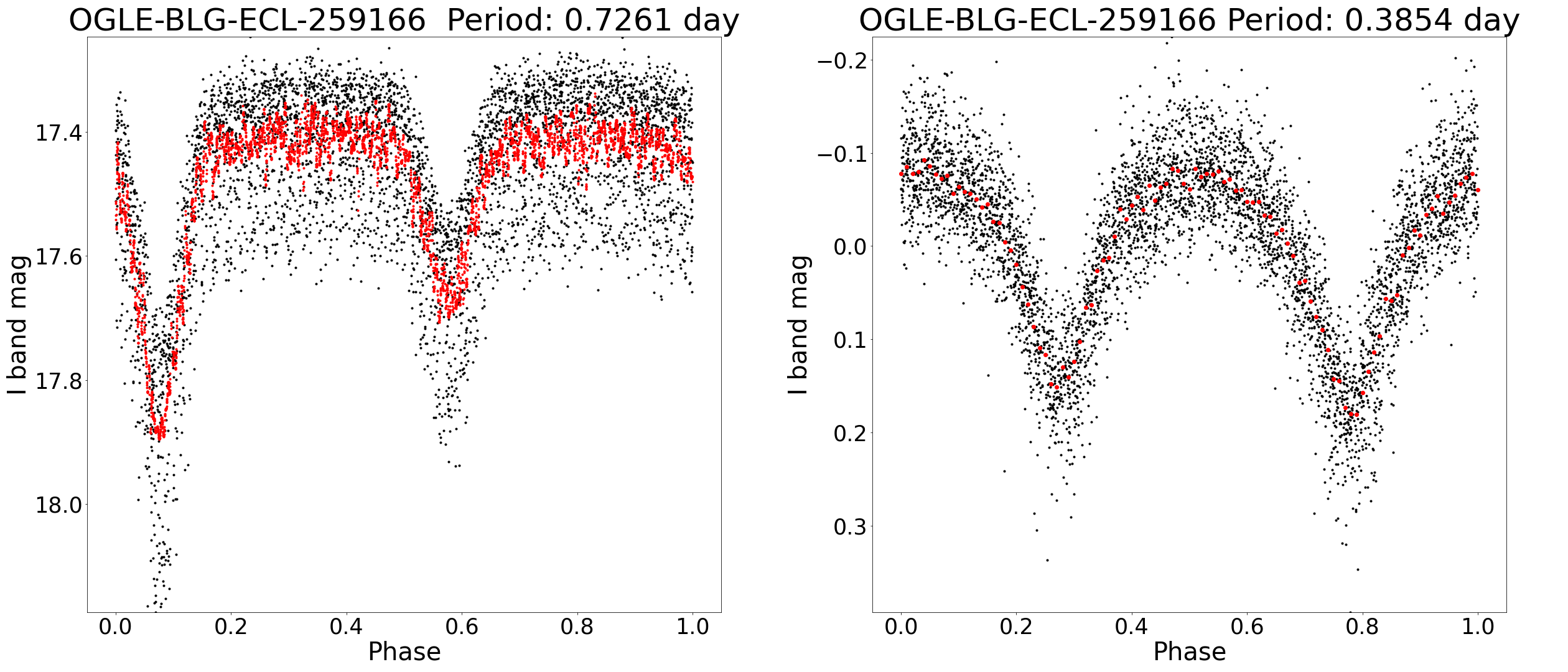}
    \includegraphics[width=\columnwidth]{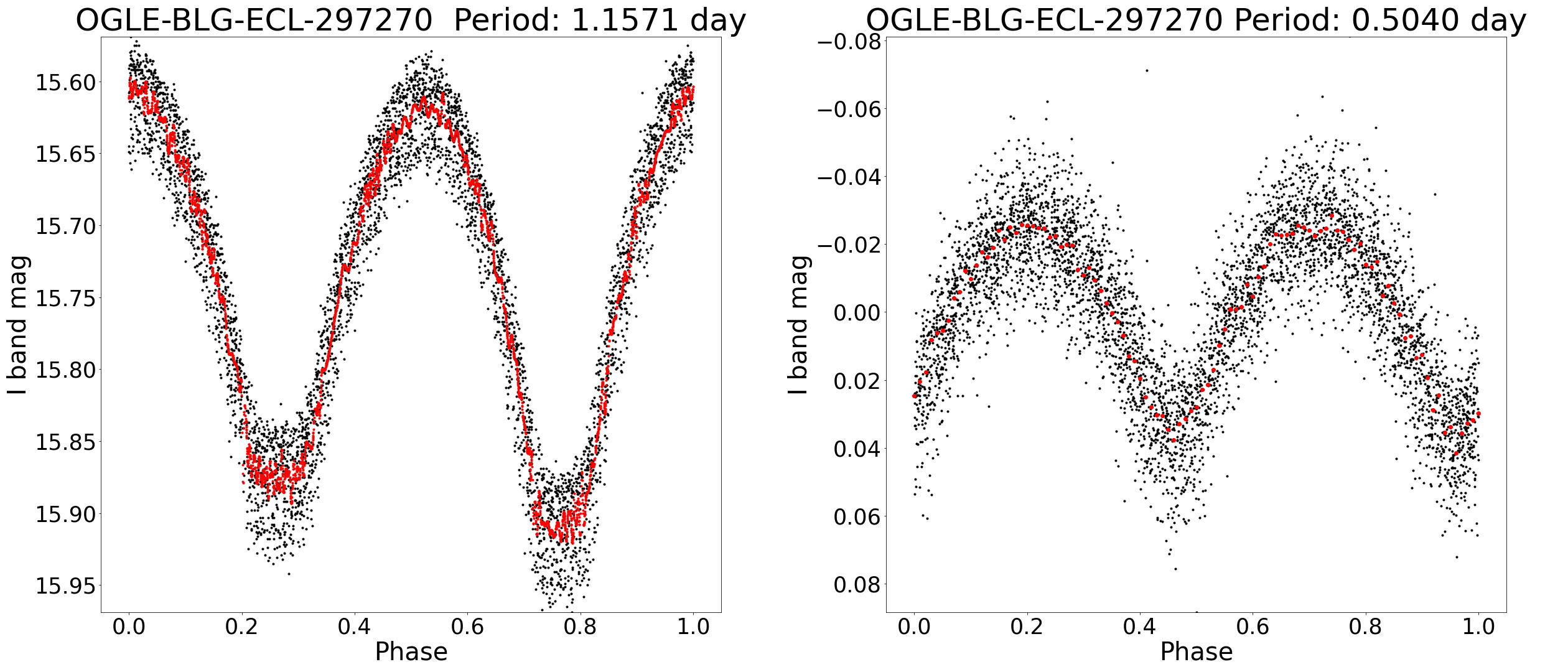}
    \includegraphics[width=\columnwidth]{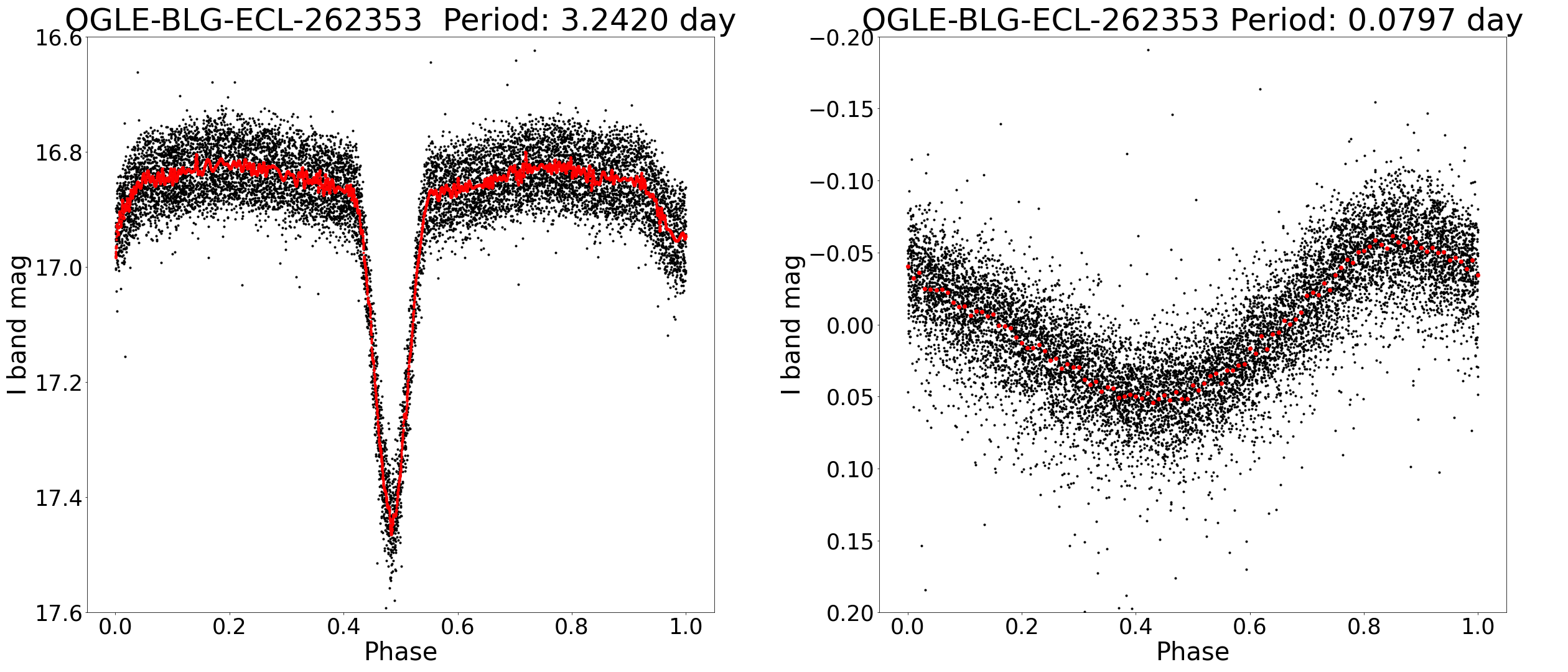}
    \caption{LCs of our candidate double EB systems and the triple candidate system with a $\delta$ Scuti variable (bottom right). The first and third columns correspond to the foreground EBs, and the second and fourth columns are the background variable stars.
    For each system, the LCs are marked with black dots, while the FBLCs are shown with red dots.}
    \label{fig:all_double_EB_lc}
\end{figure*}

As a result, we found five quadruple system candidates.
Here, we present the ETVs of each system, including OGLE-BLG-ECL-297270 and its orbital parameters, which was previously identified as a triple system by \cite{Hajdu2019}.
The list of quadruple candidates and their orbital parameters are presented in Table \ref{tab:LTTE}. The ETVs of the systems are shown in Fig.~\ref{fig:all_double_EB_oc}, where we present the LTTE model (black line) fitted to the ETVs of the EBs (red and blue marks).
The LCs of the foreground and background EBs are presented in Fig.~\ref{fig:all_double_EB_lc} with black dots, while the FBLCs are marked with red dots.

\renewcommand{\arraystretch}{1.3}
\begin{table*}
    \centering
    \caption{Orbital elements of quadruple systems from LTTE solutions.
    $P_\textup{EB}$ and $P_2$ stand for the periods of the EBs and the outer orbital period based on the LTTE solution, respectively. The $a_{\textup{EB}}\sin i_2$ is the projected semi-major axis of the LTTE orbit, while $e_2$, $\tau_2$, and $\omega_2$ stand for the eccentricity, the time of periastron passage, and the argument of periastron, respectively. The last column presents the calculated mass function of the tertiary component derived from the LTTE solutions as described in Sect. \ref{method_lte}. The identification of AB refers to the OGLE EB and CD the newly found `background' EB.}
    \label{tab:LTTE}
    \begin{tabular}{c c c c c c c c c }
\hline
ID &  $P_{\textup{EB}}$ &  $P_2$ & $a_\mathrm{EB}\cdot\sin(i_2)$ & $e_2$ & $\tau_2$ & $\omega_2$ & $f(m)$ \\
 & [days] & [days]  & [$R_{\odot}$] &  & [days] & [\degree] & [$M_{\odot}$]\\ \hline
OGLE-BLG-ECL-$130692_{\textup{AB}}$  & $0.419959$ & \multirow{2}{*}{$2530_{-113}^{+143}$} & $450_{-43}^{+70}$ & \multirow{2}{*}{$0.55_{-0.01}^{+0.1}$} &
\multirow{2}{*}{$5227_{-30}^{+34}$} & \multirow{2}{*}{$149_{-5}^{+5}$} &  0.19 \\[0.2cm]
 $130692_{\textup{CD}}$ & $0.5810375$ & & $384_{-45}^{+60}$ & & & &  0.12 \\  \hline

OGLE-BLG-ECL-$145467_{\textup{AB}}$  & $3.304929$ & \multirow{2}{*}{$1485_{-34}^{+48}$} & $564_{-19}^{+24}$ & \multirow{2}{*}{$0.44_{-0.04}^{+0.04}$} &
\multirow{2}{*}{$6014_{-13}^{+12}$} & \multirow{2}{*}{$158_{-4}^{+4}$} & 1.08 \\[0.2cm]
$145467_{\textup{CD}}$ & $4.909654$ & & $586_{-26}^{29}$ &  & & &  1.21 \\ \hline

OGLE-BLG-ECL-$246476_{\textup{AB}}$  & $2.641367$ & \multirow{2}{*}{$1624_{-16}^{+18}$} & $146_{-3}^{+3}$ & \multirow{2}{*}{$0.22_{-0.04}^{+0.04}$} &
\multirow{2}{*}{$5760_{-38}^{+45}$} & \multirow{2}{*}{$200_{-8}^{+8}$} & 0.02 \\[0.2cm]
$246476_{\textup{CD}}$ & $0.411444$ & & $145_{-3}^{+3}$ & & & &   0.02 \\ \hline

OGLE-BLG-ECL-$259166_{\textup{AB}}$  & $0.726122$ & \multirow{2}{*}{$942_{-12}^{+12}$} & $222^{+12}_{-10}$ & \multirow{2}{*}{$0.37^{+0.05}_{-0.05}$} &
\multirow{2}{*}{$5524^{+29}_{-30}$} & \multirow{2}{*}{$166_{-9}^{+9}$} & 0.16 \\[0.2cm] 
$259166_{\textup{CD}}$ & $0.385391$ & & $223^{+10}_{-9}$ & & & &   0.17 \\ \hline

OGLE-BLG-ECL-$297270_{\textup{AB}}$  & $1.157091$ & \multirow{2}{*}{$1173_{-1}^{+1}$} & $351^{+3}_{-3}$ & \multirow{2}{*}{$0.27_{-0.01}^{+0.01}$} &
\multirow{2}{*}{$3796_{-9}^{+9}$} & \multirow{2}{*}{$67^{+3}_{-3}$} & 0.42 \\[0.2cm] 
$297270_{\textup{CD}}$ & $0.503959$ & & $361_{-7}^{+6}$ & & & &   0.45 \\ \hline
    \end{tabular}
\end{table*}

\paragraph{OGLE-BLG-ECL-130692}
This system consists of a W UMa and $\beta$ Lyrae binary. Since the OGLE-IV observations do not cover a whole outer orbital cycle, we complemented the dataset with observations made by OGLE-III.
From the ETV analysis we found that this system has the longest outer orbital period and the largest eccentricity.

\paragraph{OGLE-BLG-ECL-145467}
This system was first discovered by \cite{Zasche2019} and was published with slightly different parameters compared to our results in Table \ref{tab:LTTE}.In addition to the OGLE data, which we used here, they collected observations with the Danish 1.54\,m telescope, which likely influenced their results. As far as our results are concerned, it is an eccentric system with two Algol-type EBs. 

\paragraph{OGLE-BLG-ECL-246576}
The ETV curve of this multiple system has the worst phase coverage, but the relative error of the outer period is relativity low. Of the newly identified quadruple systems, this one has the lowest outer eccentricity. This system consists of an Algol and $\beta$ Lyrae-type binary star.
Interestingly, the difference between the periods of the two EBs is the greatest ($2.435667^d$) here.

\paragraph{OGLE-BLG-ECL-259166}
This system has the shortest outer period ($P_2 = 942^d$) from our sample, almost two whole cycles appear in its $O-C$ diagram in Fig.~\ref{fig:all_double_EB_oc}, considering only OGLE-IV data.

\paragraph{OGLE-BLG-ECL-297270} 
It is the only candidate whose outer orbital period could be determined with the accuracy of 1 day since observations are available from the previous campaign (OGLE-III) and it has a relatively short period ($P_2 \sim 3.2$ yr). Four cycles are present in the $O-C$ diagram of the system, which is the bottom left panel of Fig.~\ref{fig:all_double_EB_oc}, which also includes the OGLE-III data. This system consists of a $\beta$ Lyrae and W~UMa-type binary as it can be seen in Fig.~\ref{fig:all_double_EB_lc}.

\subsection{Multiple systems from eclipse and pulsation timings}
\begin{table*}
    \centering
    \caption{Orbital elements of OGLE-BLG-ECL-262353. The meanings of the columns are the same as in Table \ref{tab:LTTE}.}
    \label{tab:LTTE_dsct}
    \begin{tabular}{c c c c c c c c c }
\hline
ID &  $P_{\textup{EB}}$ &  $P_2$ & $a_\mathrm{EB}\cdot\sin(i_2)$ & $e_2$ & $\tau_2$ & $\omega_2$ & $f(m)$ \\
 & [days] & [days]  & [$R_{\odot}$] &  & [days] & [\degree] & [$M_{\odot}$]\\ \hline
\multirow{2}{*}{OGLE-BLG-ECL-262353} & $3.242008$ & \multirow{2}{*}{$1249_{-4}^{+4}$} &  \multirow{2}{*}{$151_{-2}^{+2}$} &  \multirow{2}{*}{$0.19_{-0.02}^{+0.02}$} &  \multirow{2}{*}{$5837_{-19}^{+19}$} &  \multirow{2}{*}{$277_{-5}^{+5}$} & \multirow{2}{*}{0.03} \\ [0.2cm]
& $0.079747$ \\ \hline
    \end{tabular}
\end{table*}

As mentioned earlier, we prepared the $O-C$ diagrams of the newly found $\delta$ Scuti stars. In one case we found that the $\delta$ Scuti star and the examined EB show similar alternations in their phases. This is the OGLE-BLG-ECL-262353 system, whose $O-C$ diagram is presented in the bottom right corner of Fig.~\ref{fig:all_double_EB_oc}. For better visual transparency we subtracted $0.05^d$ from the $O-C$ diagram of the $\delta$ Scuti variable.
The matching trend suggests that the $\delta$ Scuti is a component of the close EB and there is also a distant third component in the system, resulting in the same variations on both diagrams. The outer orbital parameters of the system are listed in Table \ref{tab:LTTE_dsct}.

The OGLE team identified more than 10,000 $\delta$ Scuti stars and in some cases discovered eclipses in the LC of the pulsating variable \citep{OGLE_dsct_2020}.
We created the $O-C$ diagrams of these $\delta$ Scuti and EB stars after disentangling their LCs to examine whether they are connected like the one we had found. 
In one case the $O-C$ diagram of a $\delta$ Scuti (OGLE-BLG-DSCT-06913) showed quasi-sinusoidal variation. This object is also classified as OGLE-BLG-ELL-018626 by \citet{OGLE_BLG}. However, the EB signal identified in this system has a much noisier $O-C$ diagram, which makes it impossible to determine the relationship between the two variables.

\section{Summary and conclusions} \label{sec:summary}

In this paper we present an alternative and fast method for searching for periodic variations in residuals of EB LCs.

First, we chose an appropriate sample based on the characteristics of the dataset and selected the EB systems that were suitable for our analysis. Then, we subtracted the signals of the EBs and searched for periodicities in the residual LCs using PDM and LS methods. In the following step, we selected our candidates and classified them via a visual inspection of their LCs, with the use of Fourier parameters and with an image-based machine-learning classifier. We validated our method through tests with artificial LCs as well.

As a result, we find 354 systems that have significant periodic variations in their residual LCs. Of them, 62 are caused by already known blended variables, but in most cases (292) we find a new variable measured together with the EB.

We investigated whether the new variables are physically connected to EBs via eclipse timing and LTTE modelling. We identify four new doubly EB systems and one that was discovered earlier by \cite{Zasche2019}. Furthermore, we find a multiple system where one component of the close binary is a $\delta$ Scuti, whose $O-C$ diagram shows the same variations as the $O-C$ of the EB, indicating a distant third component.

In conclusion, we emphasise the importance of handling blends precisely, especially in dense stellar fields like the Bulge.
We note that there are many more detections in the long-period section, where different variables (e.g. Cepheids, spotted stars, etc.) are located. This dataset awaits further analysis.

\begin{acknowledgements}
We acknowledge with thanks the OGLE survey for collecting these long datasets. 
We thank the referee for their helpful comments.
R.~Z.~\'{A}d\'{a}m thanks the financial support provided by the undergraduate research assistant program of Konkoly Observatory.
This project was supported by the KKP-137523 `SeismoLab' \'Elvonal grant of the Hungarian Research, Development and Innovation Office (NKFIH) and by the Lend\"ulet Program  of the Hungarian Academy of Sciences under project No. LP2018-7. This research has made use of NASA’s Astrophysics Data System.
\end{acknowledgements}

\section*{Data Availability}
The photometry data for OGLE-IV EBs towards the Galactic Bulge is publicly available and can be downloaded from \url{http://www.astrouw.edu.pl/ogle/ogle4/OCVS/blg/ecl/}. Detected periods and disentangled LCs (as in Table \ref{table:lc_data}) are available at the CDS via anonymous ftp to cdsarc.cds.unistra.fr (130.79.128.5) or via \url{https://cdsarc.cds.unistra.fr/cgi-bin/qcat?J/A+A/}.

\begin{table}
\caption{Example table of the residual LC of OGLE-BLG-ECL-032146, whose data are available with all the other catalogued variable stars at the CDS.}
    \centering
    \begin{tabular}{l r r}
Time & Brightness & Error\\
(HJD$-2450000$) & [mag] & [mag] \\ \hline
5265.81251 & -0.036 & 0.018 \\
5266.79609 & -0.004 & 0.02 \\
5267.78687 & -0.020 & 0.019 \\
5268.8263 & -0.007 & 0.019 \\
5269.77248 & 0.007 & 0.02 \\
5270.76156 & 0.007 & 0.023 \\
5271.84391 & -0.037 & 0.02 \\
5272.75785 & -0.009 & 0.031 \\
5273.81683 & 0.004 & 0.018 \\
5274.78262 & -0.008 & 0.021 \\
\multicolumn{1}{c}{...} & ... & ... \\ \hline
    \end{tabular}
    \label{table:lc_data}
\end{table}

\bibliographystyle{aa}
\bibliography{A_cikk}

\begin{appendix}

\section{Figures of classification} \label{sec:app_class_figures}
\begin{figure*}[htb!]
    \centering
    \includegraphics[width=0.93\linewidth]{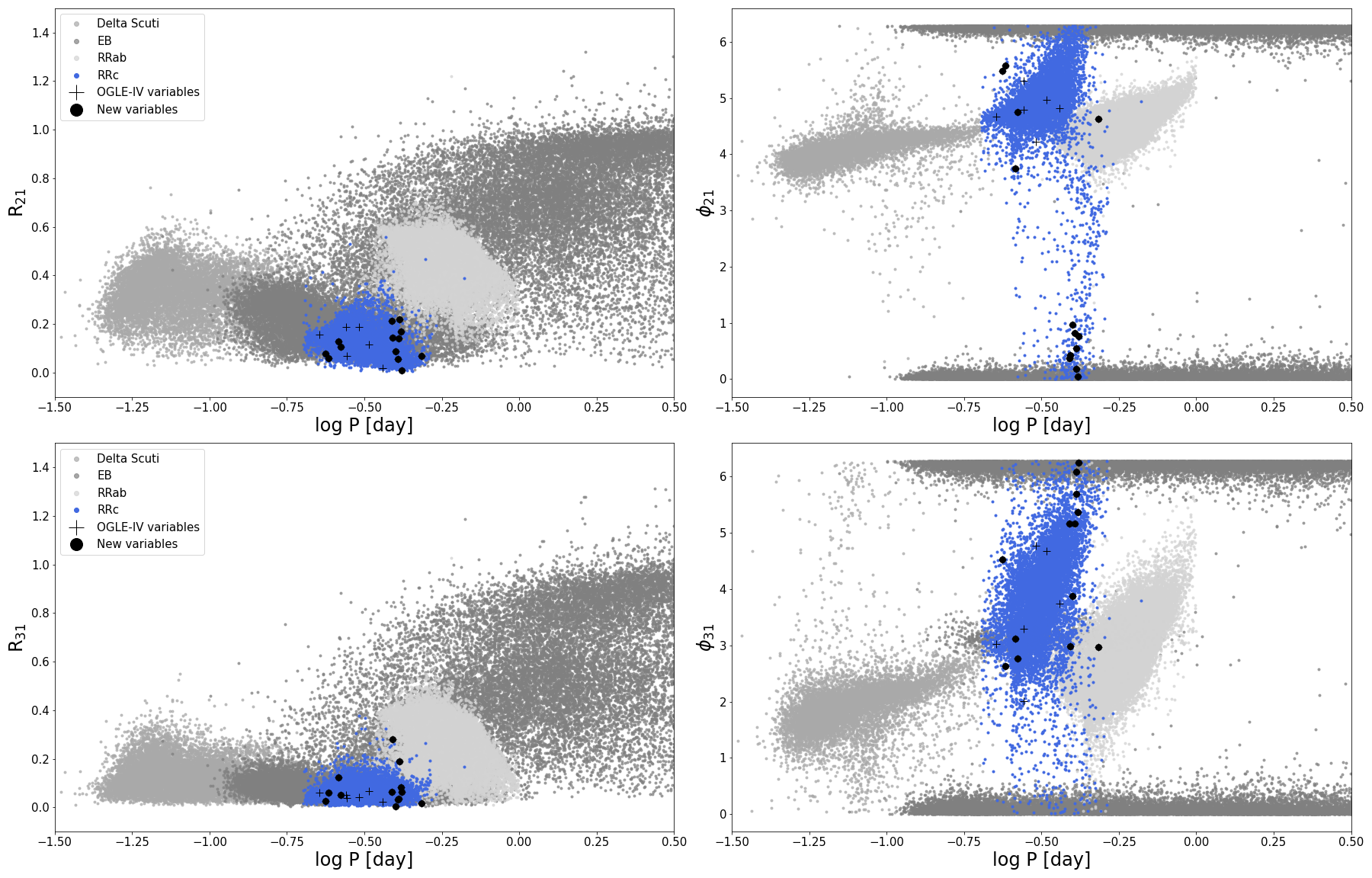}
    \caption{Example of the plots that we used to classify the pulsators based on their relative Fourier parameters, in this case RRc variables (blue).
    The background is composed of the known OGLE variable stars (except the Cepheids), as in Fig. \ref{fig:fourier_all} but now coloured in shades of grey. The dots denote the newly found variables. The crosses mark the ones that turned out to be the results of contamination by known close variables (listed in Table \ref{table:contaminated}). }
    \label{fig:fourier_rrc}
\end{figure*}

\begin{figure*}[hbt!]
    \centering
    \includegraphics[width=0.93\linewidth]{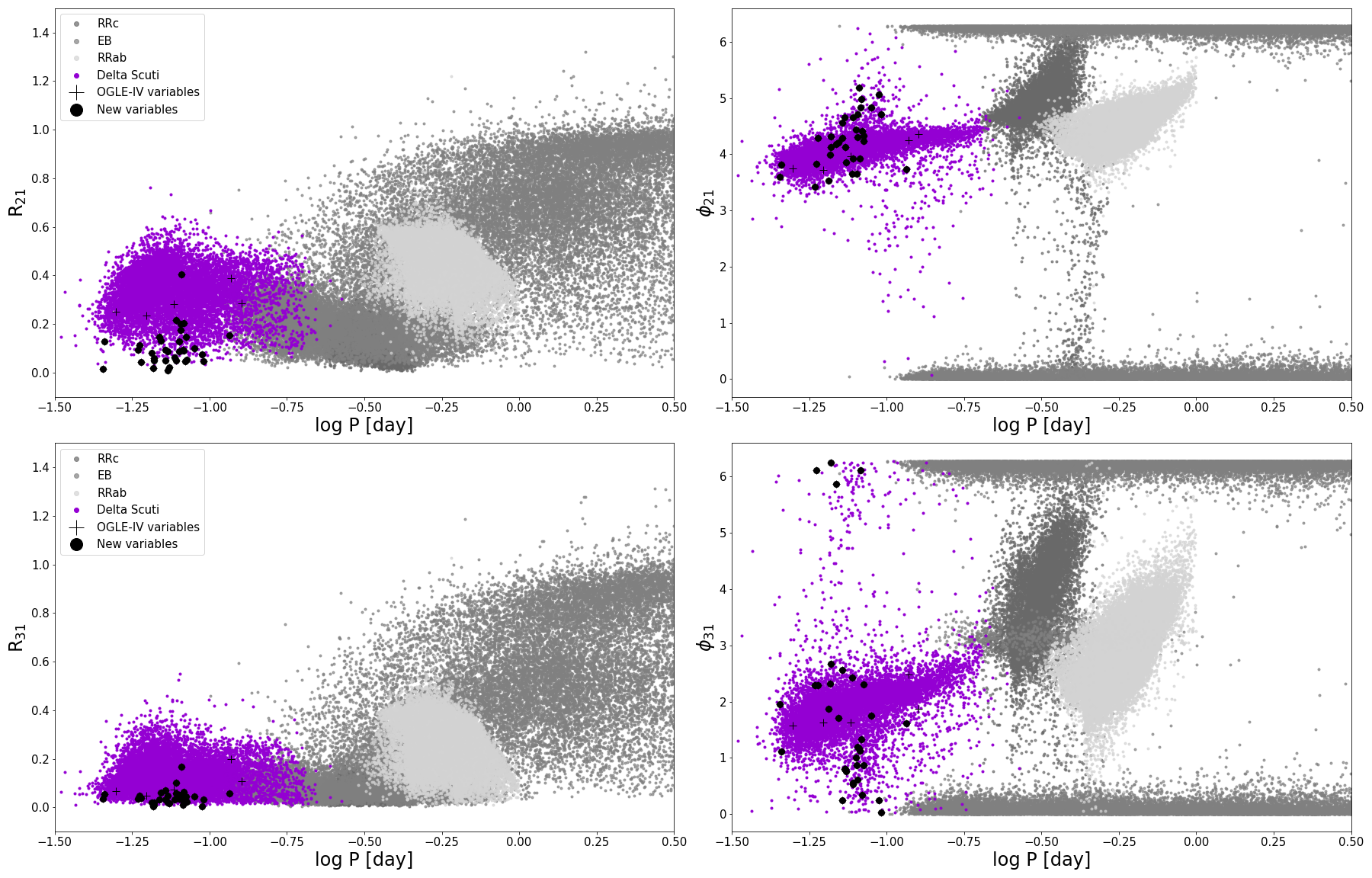}
    \caption{Same as Fig. \ref{fig:fourier_rrc}, but for $\delta$ Scuti variables (purple).}
    \label{fig:fourier_dsct}
\end{figure*}

\begin{figure*}[hbt!]
    \centering
    \includegraphics[width=0.93\linewidth]{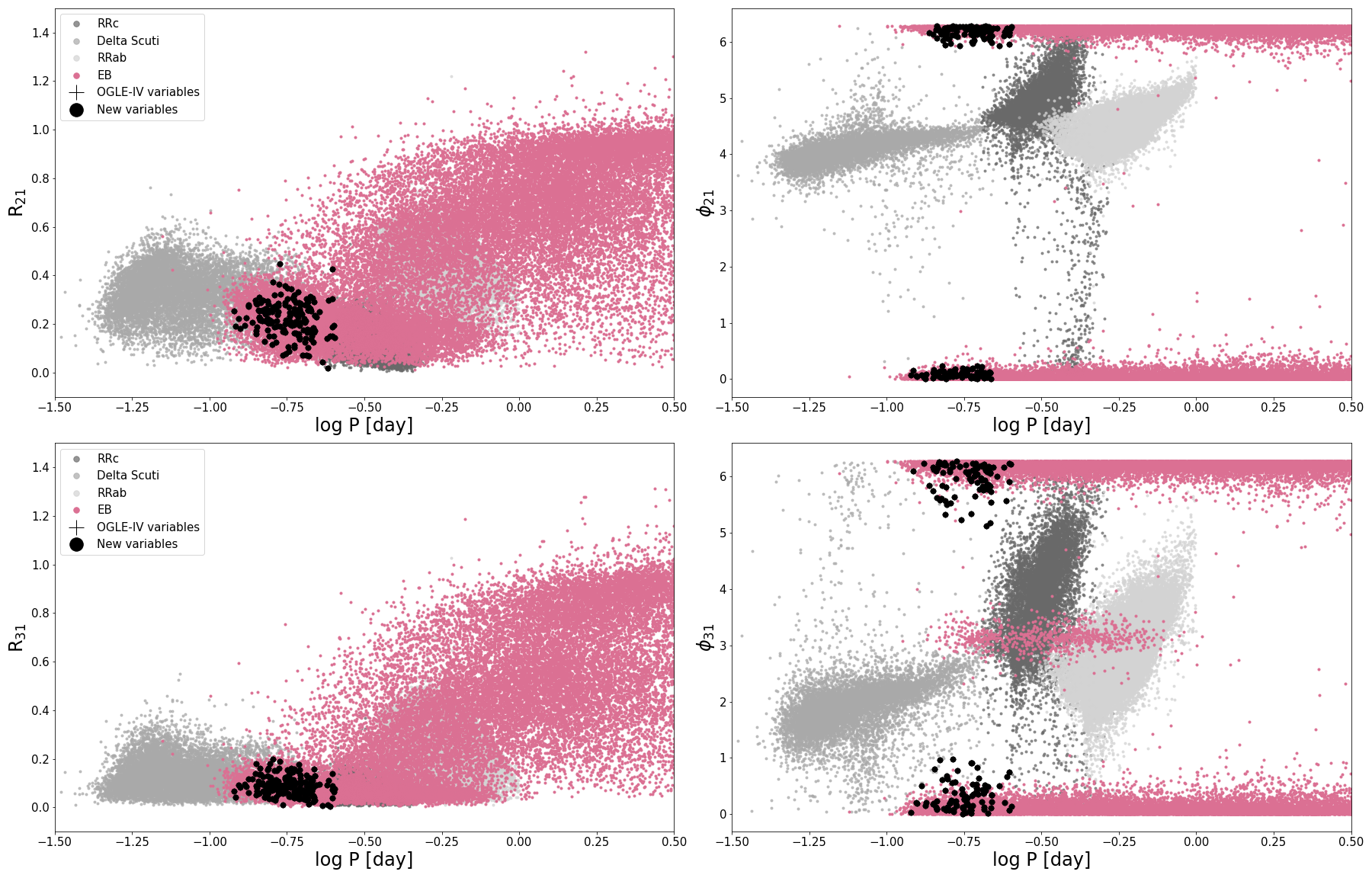}
    \caption{Same as Fig. \ref{fig:fourier_rrc}, but for W UMa variables (light pink).}
    \label{fig:fourier_wuma}
\end{figure*}

In this section we present the auxiliary figures (Figs. \ref{fig:fourier_rrc}, \ref{fig:fourier_dsct}, and \ref{fig:fourier_wuma}) we made for the classification according to relative Fourier parameters. Instead of presenting a colourful image like in Fig.~\ref{fig:fourier_all}, here we colour only the type whose new objects we would like to identify and display the others with shades of grey.

\section{Background variables}

As mentioned in Section \ref{sec:contamination}, we examined whether our detections were indeed new variables. Here, we present the whole table where we collected the eclipsing binaries whose light curves are contaminated by a known close variable.

\vspace{0.5cm}
\renewcommand{\arraystretch}{1.2}
\tablefirsthead{\toprule OGLE ID of EB & \multicolumn{1}{c}{OGLE ID of BV} \\ \midrule} 
\topcaption{Full list of the EB systems with signal of known background variable (BV).} \label{table:cont_full}
\tablehead{%
\multicolumn{2}{l}{\bfseries Table B.1. continued.} \\
\toprule
OGLE ID of EB & \multicolumn{1}{c}{OGLE ID of BV} \\ \midrule}
\begin{supertabular}{c|l}
OGLE-BLG-ECL-043403 & OGLE-BLG-RRLYR-01193 \\
OGLE-BLG-ECL-050415 & OGLE-BLG-DSCT-01176 \\
OGLE-BLG-ECL-140255 & OGLE-BLG-RRLYR-04853 \\
OGLE-BLG-ECL-162653 & OGLE-BLG-RRLYR-31535 \\
OGLE-BLG-ECL-164054 & OGLE-BLG-DSCT-03914 \\
OGLE-BLG-ECL-166633 & OGLE-BLG-RRLYR-31656 \\
OGLE-BLG-ECL-168002 & OGLE-BLG-ECL-168013 \\
OGLE-BLG-ECL-168013 & OGLE-BLG-ECL-168002 \\
OGLE-BLG-ECL-169301 & OGLE-BLG-RRLYR-31721 \\
OGLE-BLG-ECL-169385 & OGLE-BLG-ECL-169395 \\
OGLE-BLG-ECL-169395 & OGLE-BLG-ECL-169385 \\
OGLE-BLG-ECL-172630 & OGLE-BLG-RRLYR-06807 \\
OGLE-BLG-ECL-190197 & OGLE-BLG-RRLYR-07679 \\
OGLE-BLG-ECL-204060 & OGLE-BLG-ECL-204092 \\
OGLE-BLG-ECL-204092 & OGLE-BLG-ECL-204060 \\
OGLE-BLG-ECL-210139 & OGLE-BLG-RRLYR-08718 \\
OGLE-BLG-ECL-210337 & OGLE-BLG-RRLYR-08725 \\
OGLE-BLG-ECL-211367 & OGLE-BLG-ECL-211374 \\
OGLE-BLG-ECL-211374 & OGLE-BLG-ECL-211367 \\
OGLE-BLG-ECL-223081 & OGLE-BLG-RRLYR-09417 \\
OGLE-BLG-ECL-224671 & OGLE-BLG-RRLYR-32979 \\
OGLE-BLG-ECL-225293 & OGLE-BLG-ECL-225302 \\
OGLE-BLG-ECL-225302 & OGLE-BLG-ECL-225293 \\
OGLE-BLG-ECL-227744 & OGLE-BLG-ECL-227750 \\
OGLE-BLG-ECL-227750 & OGLE-BLG-ECL-227744 \\
OGLE-BLG-ECL-232049 & OGLE-BLG-ECL-232059 \\
OGLE-BLG-ECL-232059 & OGLE-BLG-ECL-232049 \\
OGLE-BLG-ECL-233822 & OGLE-BLG-ECL-233847 \\
OGLE-BLG-ECL-233847 & OGLE-BLG-ECL-233822 \\
OGLE-BLG-ECL-234434 & OGLE-BLG-DSCT-05682 \\
OGLE-BLG-ECL-234828 & OGLE-BLG-ECL-234844 \\
OGLE-BLG-ECL-234844 & OGLE-BLG-ECL-234828 \\
OGLE-BLG-ECL-239150 & OGLE-BLG-DSCT-05792 \\
OGLE-BLG-ECL-239517 & OGLE-BLG-RRLYR-10302 \\
OGLE-BLG-ECL-239777 & OGLE-BLG-ECL-239801 \\
OGLE-BLG-ECL-239801 & OGLE-BLG-ECL-239777 \\
OGLE-BLG-ECL-245466 & OGLE-BLG-ECL-245476 \\
OGLE-BLG-ECL-245476 & OGLE-BLG-ECL-245466 \\
OGLE-BLG-ECL-246036 & OGLE-BLG-ECL-246041 \\
OGLE-BLG-ECL-246041 & OGLE-BLG-ECL-246036 \\
OGLE-BLG-ECL-246473 & OGLE-BLG-ECL-246468 \\
OGLE-BLG-ECL-246468 & OGLE-BLG-ECL-246473 \\
OGLE-BLG-ECL-248809 & OGLE-BLG-RRLYR-10759 \\
OGLE-BLG-ECL-256053 & OGLE-BLG-ECL-256059 \\
OGLE-BLG-ECL-256059 & OGLE-BLG-ECL-256053 \\
OGLE-BLG-ECL-261418 & OGLE-BLG-ECL-261424 \\
OGLE-BLG-ECL-261424 & OGLE-BLG-ECL-261418 \\
OGLE-BLG-ECL-263077 & OGLE-BLG-DSCT-06348 \\
OGLE-BLG-ECL-269577 & OGLE-BLG-ECL-269580 \\
OGLE-BLG-ECL-269580 & OGLE-BLG-ECL-269577 \\
OGLE-BLG-ECL-269991 & OGLE-BLG-ECL-269995 \\
OGLE-BLG-ECL-269995 & OGLE-BLG-ECL-269991 \\
OGLE-BLG-ECL-271895 & OGLE-BLG-RRLYR-11886 \\
OGLE-BLG-ECL-279001 & OGLE-BLG-ECL-279020 \\
OGLE-BLG-ECL-279020 & OGLE-BLG-ECL-279001 \\
OGLE-BLG-ECL-280921 & OGLE-BLG-ECL-280936 \\
OGLE-BLG-ECL-280936 & OGLE-BLG-ECL-280921 \\
OGLE-BLG-ECL-285148 & OGLE-BLG-ECL-285159 \\
OGLE-BLG-ECL-285159 & OGLE-BLG-ECL-285148 \\
OGLE-BLG-ECL-285403 & OGLE-BLG-ECL-285426 \\
OGLE-BLG-ECL-285426 & OGLE-BLG-ECL-285403 \\
OGLE-BLG-ECL-297739 & OGLE-BLG-RRLYR-12948 \\
\end{supertabular}
\vspace{1cm}

\clearpage
\newpage
\noindent Through our study we identified 292 new variable stars in the `backgrounds' of OGLE-IV EBs.  In Table \ref{table:var_full} we present the full catalogue of them, listing the identifiers and periods of the OGLE EBs, as well as the periods and types of the detected `background' variables.

\vspace{0.5cm}
\tablefirsthead{\toprule ID & \multicolumn{1}{c}{$P_{\textup{EB}}$} & \multicolumn{1}{c}{$P_{\textup{BG}}$} & \multicolumn{1}{c}{Type} \\ 
(OGLE-BLG-ECL-) & \multicolumn{1}{c}{[day]} & \multicolumn{1}{c}{[day]} & \\\midrule}
\tablecaption{Full list of EB systems with a new variable in the background. This table is also available at the CDS.} \label{table:var_full}
\tablehead{\multicolumn{4}{l}{{\bfseries Table B.2. continued from previous column.}} \\
\toprule ID & \multicolumn{1}{c}{$P_{\textup{EB}}$} & \multicolumn{1}{c}{$P_{\textup{BG}}$} & \multicolumn{1}{c}{Type} \\ 
(OGLE-BLG-ECL-) & \multicolumn{1}{c}{[day]} & \multicolumn{1}{c}{[day]} & \\\midrule}
\begin{supertabular}{llll}
028118 & 0.715801 & 0.281800 & W UMa \\
032146 & 1.865528 & 0.077299 & $\delta$ Scuti \\
033287 & 0.321652 & 0.350422 & W UMa \\
036387 & 0.843114 & 0.478542 & W UMa \\
036660 & 1.980759 & 0.398770 & W UMa \\
038325 & 0.664686 & 0.347723 & W UMa \\
039882 & 2.363820 & 0.436686 & W UMa \\
039897 & 3.786951 & 0.068672 & $\delta$ Scuti \\
085710 & 0.926231 & 0.274207 & W UMa \\
111212 & 1.021946 & 0.376466 & W UMa \\
111324 & 0.250133 & 1.154810 & Algol \\
117283 & 0.374526 & 0.410105 & W UMa \\
120853 & 0.647914 & 5.681075 & Algol \\
121034 & 0.503087 & 0.328665 & W UMa \\
123507 & 1.016974 & 2.948768 & Algol \\
124829 & 6.985597 & 3.139287 & Algol \\
124975 & 2.867577 & 2.040574 & Algol \\
129019 & 0.591868 & 0.412415 & W UMa \\
134784 & 0.990926 & 0.367435 & W UMa \\
134943 & 0.802096 & 0.351285 & W UMa \\
136001 & 5.335628 & 0.908798 & Algol \\
136247 & 0.762326 & 1.152883 & Algol \\
136305 & 0.682184 & 0.416493 & RRc \\
138459 & 1.541225 & 6.641701 & Algol \\
139222 & 0.457451 & 0.481934 & W UMa \\
140911 & 4.625392 & 0.089122 & $\delta$ Scuti \\
141382 & 0.959037 & 0.409633 & W UMa \\
141909 & 1.166872 & 0.388954 & RRc \\
145377 & 3.091635 & 0.307654 & W UMa \\
145467 & 3.304929 & 4.909654 & Algol \\
145640 & 4.959043 & 0.080702 & $\delta$ Scuti \\
145963 & 2.161134 & 0.077852 & $\delta$ Scuti \\
148466 & 4.978612 & 0.069697 & $\delta$ Scuti \\
151177 & 4.600011 & 2.295456 & Algol \\
151421 & 0.492524 & 0.378721 & W UMa \\
151687 & 0.873994 & 0.307138 & W UMa \\
151906 & 0.564080 & 0.293391 & W UMa \\
152990 & 5.376781 & 0.094626 & $\delta$ Scuti \\
153633 & 1.097582 & 0.362195 & W UMa \\
154538 & 1.471585 & 0.485966 & W UMa \\
155419 & 0.717134 & 1.588791 & Algol \\
156530 & 0.559385 & 0.437271 & W UMa \\
156651 & 0.409973 & 0.302892 & W UMa \\
156811 & 0.565451 & 0.462250 & W UMa \\
157170 & 0.470791 & 0.291062 & W UMa \\
157834 & 0.409131 & 0.409591 & RRc \\
158241 & 0.393769 & 0.309049 & W UMa \\
158310 & 0.466152 & 0.732631 %0.7326305
& Algol \\
159690 & 1.440827 & 0.073376 & $\delta$ Scuti \\
160180 & 0.915273 & 0.422605 & W UMa \\
162527 & 0.469210 & 0.419149 & W UMa \\
162682 & 30.509576 & 0.497508 & W UMa \\
162984 & 6.492355 & 0.073097 & $\delta$ Scuti \\
163992 & 1.038155 & 0.668439 & W UMa \\
164230 & 1.289267 & 2.990437 & Algol \\
165251 & 0.664308 & 0.386396 & W UMa \\
165412 & 0.347736 & 3.632327 & Algol \\
165625 & 0.312758 & 0.264897 & RRc \\
165896 & 0.516570 & 1.795378 & Algol \\
166629 & 0.514403 & 0.960745 & Algol \\
166686 & 0.498144 & 1.363588 & Algol \\
166724 & 0.479567 & 0.313125 & W UMa \\
166838 & 1.127588 & 0.322343 & W UMa \\
167334 & 4.410296 & 0.090632 & W UMa \\
170025 & 1.589483 & 0.412418 & W UMa \\
170305 & 0.295228 & 0.352196 & W UMa \\
170306 & 1.503033 & 0.081990 & $\delta$ Scuti \\
171575 & 0.571010 & 0.236434 & RRc \\
172418 & 1.107471 & 3.119255 & Algol \\
172918 & 0.733702 & 0.453118 & W UMa \\
174119 & 0.762574 & 3.287503 & $\beta$ Lyrae \\
174972 & 1.807395 & 0.415665 & W UMa \\
175363 & 5.836193 & 0.332606 & W UMa \\
175389 & 9.962233 & 0.369661 & W UMa \\
175451 & 1.038538 & 0.407096 & W UMa \\
175866 & 32.155974 & 0.364633 & W UMa \\
176073 & 0.797012 & 0.351255 & W UMa \\
176216 & 0.535362 & 0.344023 & W UMa \\
176470 & 0.805849 & 4.331099 & Algol \\
178352 & 0.693854 & 0.115642 & $\delta$ Scuti \\
179134 & 0.404242 & 0.434915 & W UMa \\
180524 & 0.297821 & 0.433165 & W UMa \\
180559 & 0.865819 & 0.260299 & RRc \\
182502 & 6.172291 & 0.081234 & $\delta$ Scuti \\
183527 & 0.336450 & 0.415538 & W UMa \\
183649 & 2.737795 & 0.391084 & RRc \\
183765 & 0.429345 & 0.319393 & W UMa \\
184951 & 0.790206 & 0.389627 & W UMa \\
186032 & 1.482376 & 0.309960 & W UMa \\
186346 & 0.928168 & 0.497266 & W UMa \\
186604 & 6.629573 & 0.325722 & W UMa \\
187122 & 0.774036 & 0.331701 & W UMa \\
189104 & 0.459364 & 0.425778 & W UMa \\
189844 & 1.449876 & 0.302077 & W UMa \\
191453 & 0.785358 & 1.211576 & W UMa \\
191959 & 1.352531 & 0.064894 & $\delta$ Scuti \\
192547 & 0.596436 & 0.416548 & W UMa \\
192582 & 2.455050 & 0.073907 & $\delta$ Scuti \\
193282 & 0.615897 & 0.330384 & W UMa \\
193542 & 10.864226 & 39.331636 & W UMa \\
194314 & 0.360090 & 0.376154 & W UMa \\
195851 & 7.321952 & 0.374819 & W UMa \\
196141 & 0.222942 & 0.371742 & W UMa \\
197037 & 1.041253 & 0.336643 & W UMa \\
197453 & 0.667352 & 2.597924 & Algol \\
197801 & 0.400963 & 0.320494 & W UMa \\
197897 & 0.366223 & 0.300890 & W UMa \\
198696 & 3.042769 & 0.066207 & $\delta$ Scuti \\
198783 & 0.507892 & 0.391994 & W UMa \\
199033 & 0.765743 & 0.417614 & W UMa \\
200022 & 0.350380 & 0.423114 & W UMa \\
200310 & 0.360795 & 0.333717 & W UMa \\
200402 & 0.457504 & 0.402577 & W UMa \\
202842 & 0.565191 & 0.435189 & W UMa \\
204851 & 0.574790 & 0.430802 & W UMa \\
205211 & 0.856118 & 0.263132 & W UMa \\
205414 & 1.099712 & 0.355854 & W UMa \\
205524 & 1.854937 & 0.389708 & W UMa \\
206767 & 3.105015 & 0.060025 & $\delta$ Scuti \\
207049 & 0.468535 & 0.316510 & $\beta$ Lyrae \\
207289 & 0.984431 & 0.360314 & W UMa \\
207490 & 4.028993 & 0.077778 & $\delta$ Scuti \\
207504 & 0.414866 & 0.403268 & W UMa \\
207581 & 0.369578 & 0.334869 & W UMa \\
208337 & 4.268947 & 0.991300 & $\beta$ Lyrae \\
209383 & 0.388243 & 0.282322 & W UMa \\
209728 & 0.322124 & 0.336769 & W UMa \\
210254 & 1.173553 & 0.373381 & W UMa \\
211260 & 0.624758 & 0.389399 & W UMa \\
212142 & 0.466858 & 0.313969 & W UMa \\
212994 & 0.690829 & 0.317821 & W UMa \\
213783 & 0.415471 & 0.379676 & W UMa \\
213786 & 0.379677 & 0.415471 & W UMa \\
213949 & 0.691431 & 2.115856 & Algol \\
214088 & 0.429557 & 0.406981 & W UMa \\
214815 & 6.152946 & 0.065780 & $\delta$ Scuti \\
215916 & 0.440753 & 0.676875 & $\beta$ Lyrae \\
216018 & 0.362244 & 0.432300 & W UMa \\
216324 & 0.643459 & 0.392652 & W UMa \\
217014 & 0.500621 & 0.310388 & W UMa \\
217054 & 1.006902 & 0.498366 & W UMa \\
218391 & 2.074209 & 0.045654 & $\delta$ Scuti \\
218440 & 2.970689 & 0.379868 & W UMa \\
218771 & 0.398029 & 0.441988 & W UMa \\
219072 & 0.431044 & 1.163686 & Algol \\
219076 & 0.488153 & 0.296672 & W UMa \\
219778 & 0.430066 & 0.251226 & W UMa \\
219999 & 0.419220 & 0.288259 & W UMa \\
220530 & 0.552028 & 0.397619 & W UMa \\
220632 & 1.295319 & 2.578413 & $\beta$ Lyrae \\
222540 & 1.688043 & 0.289766 & W UMa \\
223171 & 0.427156 & 0.323270 & W UMa \\
224598 & 0.438941 & 1.588111 & Algol \\
224783 & 0.614930 & 0.289775 & W UMa \\
226277 & 0.747550 & 3.470842 & Algol \\
226796 & 0.843384 & 0.345039 & W UMa \\
226992 & 1.385150 & 0.505956 & W UMa \\
227207 & 0.604965 & 0.333697 & W UMa \\
227744 & 1.380442 & 0.419004 & $\beta$ Lyrae \\
227914 & 1.034848 & 0.489439 & W UMa \\
228244 & 0.673756 & 0.427940 & W UMa \\
228293 & 0.737306 & 0.405673 & W UMa \\
228484 & 0.373617 & 0.239816 & W UMa \\
228616 & 11.608892 & 0.095840 & $\delta$ Scuti \\
229158 & 0.620128 & 0.303132 & W UMa \\
229495 & 0.995846 & 1.288566 & Algol \\
229988 & 6.148337 & 11.705101 & Algol \\
231711 & 6.318493 & 0.082464 & $\delta$ Scuti \\
231963 & 0.416579 & 0.248875 & W UMa \\
232401 & 0.327329 & 6.113123 & Algol \\
233287 & 0.750751 & 10.729897 & Algol \\
233663 & 0.360344 & 0.375665 & W UMa \\
233819 & 0.600175 & 0.289247 & W UMa \\
233822 & 1.165683 & 0.389816 & W UMa \\
235127 & 9.048235 & 0.500515 & W UMa \\
235154 & 0.500514 & 9.047202 & $\beta$ Lyrae \\
235353 & 0.339321 & 0.408500 & RRc \\
235373 & 0.817006 & 0.339319 & W UMa \\
235899 & 0.418144 & 0.066041 & $\delta$ Scuti \\
235941 & 0.386342 & 0.267781 & W UMa \\
236897 & 0.461052 & 0.382470 & W UMa \\
237830 & 0.598003 & 0.319252 & W UMa \\
238138 & 0.416141 & 0.428540 & W UMa \\
238361 & 0.591129 & 0.386196 & W UMa \\
238988 & 4.040877 & 0.359176 & W UMa \\
239777 & 5.213709 & 0.717551 & W UMa \\
239827 & 0.436735 & 0.379564 & W UMa \\
240175 & 0.519637 & 0.575412 & W UMa \\
240177 & 0.575413 & 0.519634 & W UMa \\
240302 & 2.148774 & 0.545142 & $\beta$ Lyrae \\
240405 & 0.601882 & 0.375558 & W UMa \\
240562 & 0.339101 & 0.302779 & W UMa \\
240811 & 0.712463 & 0.329693 & W UMa \\
241147 & 0.380456 & 0.289784 & W UMa \\
241466 & 0.600617 & 1.315600 & Algol \\
243090 & 4.017878 & 0.071906 & $\delta$ Scuti \\
243295 & 0.557232 & 0.405677 & RRc \\
243520 & 0.515139 & 0.351293 & W UMa \\
244818 & 0.390002 & 0.308308 & W UMa \\
244880 & 4.218475 & 0.084273 & $\delta$ Scuti \\
245466 & 0.685942 & 0.450982 & W UMa \\
245709 & 0.381460 & 0.414254 & RRc \\
246026 & 0.304459 & 0.283018 & W UMa \\
246036 & 0.488428 & 0.718922 & W UMa \\
246468 & 3.117342 & 0.205621 & W UMa \\
246476 & 2.641367 & 0.411445 & W UMa \\
247628 & 3.985074 & 0.083382 & $\delta$ Scuti \\
247935 & 3.354309 & 0.059445 & $\delta$ Scuti \\
249084 & 0.277218 & 0.310097 & W UMa \\
249324 & 7.009192 & 0.082692 & $\delta$ Scuti \\
249429 & 0.398739 & 0.360530 & W UMa \\
250817 & 0.439616 & 9.251576 & Algol \\
250930 & 0.505607 & 0.327289 & W UMa \\
251418 & 0.695735 & 6.857099 & Algol \\
251606 & 0.334013 & 0.327027 & W UMa \\
251710 & 0.334663 & 0.365683 & W UMa \\
253194 & 0.561837 & 0.324601 & W UMa \\
254726 & 2.918204 & 0.058703 & $\delta$ Scuti \\
256025 & 0.358097 & 3.158500 & $\beta$ Lyrae \\
256298 & 0.510839 & 0.283396 & W UMa \\
257772 & 1.608808 & 0.499251 & $\beta$ Lyrae \\
258138 & 0.308203 & 1.174757 & Algol \\
258518 & 0.314944 & 0.483426 & RRc \\
258936 & 0.451274 & 0.423425 & W UMa \\
259120 & 0.349601 & 0.311689 & W UMa \\
259166 & 0.726122 & 0.385391 & W UMa \\
259321 & 0.446969 & 0.380738 & W UMa \\
259359 & 0.401639 & 0.259702 & W UMa \\
260224 & 0.357231 & 0.197554 & W UMa \\
260240 & 0.395107 & 0.178616 & W UMa \\
260512 & 0.421083 & 6.295878 & Algol \\
261116 & 0.347991 & 1.572827 & Algol \\
262353 & 3.242008 & 0.079747 & $\delta$ Scuti \\
262375 & 0.413706 & 0.345074 & W UMa \\
262725 & 4.371701 & 0.308752 & W UMa \\
263181 & 0.679875 & 0.285169 & W UMa \\
266182 & 0.348859 & 0.399107 & RRc \\
266487 & 0.760567 & 0.360594 & W UMa \\
266782 & 0.600283 & 0.297912 & W UMa \\
268375 & 6.636373 & 0.080472 & $\delta$ Scuti \\
269074 & 0.408812 & 0.331465 & W UMa \\
269995 & 0.656073 & 0.511003 & W UMa \\
270233 & 2.200611 & 0.431809 & W UMa \\
271630 & 2.431887 & 0.365328 & W UMa \\
272108 & 0.360299 & 0.404463 & W UMa \\
272656 & 3.385598 & 0.362125 & W UMa \\
272973 & 8.297816 & 0.387862 & W UMa \\
273617 & 2.773322 & 0.498201 & W UMa \\
273738 & 0.394975 & 0.688019 & Algol \\
273847 & 0.377821 & 0.314548 & W UMa \\
274534 & 0.416001 & 0.375685 & W UMa \\
274965 & 1.341903 & 0.387542 & W UMa \\
275618 & 0.402487 & 0.319819 & W UMa \\
276258 & 0.826182 & 0.418102 & W UMa \\
277014 & 0.435768 & 0.292178 & W UMa \\
277461 & 0.679686 & 0.373403 & W UMa \\
277539 & 0.577982 & 0.375328 & W UMa \\
277992 & 2.209717 & 0.391224 & W UMa \\
278817 & 0.864780 & 0.306217 & W UMa \\
279001 & 0.815092 & 0.241045 & W UMa \\
279436 & 4.162004 & 0.084103 & $\delta$ Scuti \\
279811 & 7.378003 & 0.297204 & W UMa \\
280398 & 0.407393 & 7.261806 & W UMa \\
280921 & 8.075665 & 0.356790 & W UMa \\
282464 & 0.851621 & 0.295531 & W UMa \\
283579 & 0.372345 & 1.961150 & Algol \\
284282 & 0.817397 & 0.442131 & $\beta$ Lyrae \\
284418 & 0.458769 & 0.266684 & W UMa \\
284896 & 0.226775 & 0.391970 & Algol \\
285403 & 1.077844 & 0.157675 & W UMa \\
285692 & 0.553283 & 2.059185 & Algol \\
285827 & 0.272541 & 0.424964 & W UMa \\
286273 & 0.256666 & 0.408718 & W UMa \\
286466 & 0.417993 & 0.413627 & W UMa \\
286630 & 0.549204 & 0.375750 & W UMa \\
288432 & 0.387878 & 3.707500 & Algol \\
289079 & 5.007585 & 13.531500 & Algol \\
289379 & 1.886366 & 0.079920 & $\delta$ Scuti \\
290826 & 7.179383 & 0.353315 & W UMa \\
292142 & 1.469409 & 0.071905 & $\delta$ Scuti \\
293399 & 0.423756 & 0.284270 & W UMa \\
293405 & 0.284271 & 0.423757 & W UMa \\
293627 & 3.354723 & 1.035652 & Algol \\
293769 & 1.504857 & 0.045191 & $\delta$ Scuti \\
294525 & 0.459517 & 0.242235 & RRc \\
296490 & 0.426569 & 0.391781 & W UMa \\
297270 & 1.157091 & 0.503959 & W UMa \\
297854 & 0.861031 & 0.349828 & W UMa \\
298128 & 0.513481 & 0.392259 & W UMa \\
298817 & 0.416672 & 0.393550 & W UMa \\
301991 & 0.353371 & 0.432885 & W UMa \\
303226 & 3.369807 & 0.243346 & W UMa \\
\end{supertabular}
\vspace{1cm}

\section{Results of the image-based classification}

To test our classification we used an image-based machine learning classification method  as well \citep{2020Szklenar,Szklenar22}, whose results are presented in Table \ref{table:class_image}. We marked the best variability type when its probability was greater than 80\%, and the two with the highest scores otherwise.

\vspace{0.5cm}
\renewcommand{\arraystretch}{1.2}
\begin{center}
\tablefirsthead{\toprule ID & Most probable \\ 
(OGLE-BLG-ECL-) & variable type \\ \midrule}
\tablecaption{The results of image-based classification. This table is also available at the CDS.} \label{table:class_image}
\tablehead{ 
\multicolumn{2}{l}{\bfseries Table C.1. continued from previous column.} \\
\toprule ID & Most probable \\ 
(OGLE-BLG-ECL-) & variable type \\ \midrule}
\begin{supertabular}{cl}
\linespread{1.5}
028118 & W UMa \\
032146 & $\delta$ Scuti \\
033287 & W UMa \\
036387 & W UMa \\
036660 & W UMa \\
038325 & W UMa \\
039882 & W UMa \\
039897 & $\delta$ Scuti \\
085710 & W UMa \\
111212 & W UMa \\
111324 & Algol \\
117283 & W UMa \\
120853 & Algol \\
121034 & W UMa \\
123507 & Algol \\
124829 & $\beta$ Lyrae / Algol \\
124975 & Algol \\
129019 & W UMa \\
134784 & W UMa \\
134943 & W UMa \\
136001 & W UMa \\
136247 & Algol \\
136305 & RRd / RRab \\
138459 & Algol / $\beta$ Lyrae \\
139222 & W UMa \\
140911 & $\delta$ Scuti \\
141382 & W UMa \\
141909 & RRd \\
145377 & W UMa \\
145467 & Algol \\
145640 & $\delta$ Scuti \\
145963 & $\delta$ Scuti \\
148466 & $\delta$ Scuti \\
151177 & $\beta$ Lyrae / Algol \\
151421 & W UMa \\
151687 & W UMa \\
151906 & W UMa \\
152990 & $\delta$ Scuti \\
153633 & W UMa \\
154538 & W UMa \\
155419 & Algol \\
156530 & W UMa \\
156651 & W UMa \\
156811 & W UMa \\
157170 & W UMa \\
157834 & W UMa \\
158241 & W UMa \\
158310 & W UMa \\
159690 & $\delta$ Scuti \\
160180 & W UMa \\
162527 & W UMa \\
162682 & W UMa \\
162984 & $\delta$ Scuti \\
163992 & W UMa / $\beta$ Lyrae \\
164230 & $\beta$ Lyrae / Algol \\
165251 & W UMa \\
165412 & Algol \\
165625 & W UMa / $\delta$ Scuti \\
165896 & $\beta$ Lyrae / Algol \\
166629 & $\beta$ Lyrae / W UMa \\
166686 & W UMa / $\beta$ Lyrae \\
166724 & W UMa \\
166838 & W UMa \\
167334 & $\delta$ Scuti \\
170025 & W UMa \\
170305 & W UMa / $\beta$ Lyrae \\
170306 & $\delta$ Scuti \\
171575 & $\delta$ Scuti  / RRd \\
172418 & Algol / $\beta$ Lyrae \\
172918 & W UMa \\
174119 & $\beta$ Lyrae \\
174972 & W UMa \\
175363 & W UMa \\
175389 & W UMa \\
175451 & W UMa \\
175866 & W UMa \\
176073 & W UMa \\
176216 & W UMa \\
176470 & Algol / $\beta$ Lyrae \\
178352 & $\delta$ Scuti \\
179134 & W UMa \\
180524 & W UMa \\
180559 & $\delta$ Scuti / RRd \\
182502 & $\delta$ Scuti \\
183527 & W UMa \\
183649 & W UMa \\
183765 & W UMa \\
184951 & W UMa \\
186032 & W UMa \\
186346 & W UMa \\
186604 & W UMa \\
187122 & W UMa \\
189104 & W UMa \\
189844 & W UMa \\
191453 & $\beta$ Lyrae / W UMa \\
191959 & $\delta$ Scuti \\
192547 & W UMa \\
192582 & $\delta$ Scuti \\
193282 & W UMa \\
193542 & $\beta$ Lyrae / W UMa \\
194314 & W UMa \\
195851 & W UMa \\
196141 & W UMa \\
197037 & W UMa \\
197453 & Algol \\
197801 & W UMa \\
197897 & W UMa \\
198696 & $\delta$ Scuti \\
198783 & W UMa \\
199033 & W UMa \\
200022 & W UMa / $\beta$ Lyrae \\
200310 & W UMa \\
200402 & W UMa \\
202842 & W UMa \\
204851 & W UMa \\
205211 & W UMa \\
205414 & W UMa \\
205524 & W UMa \\
206767 & $\delta$ Scuti \\
207049 & W UMa \\
207289 & W UMa \\
207490 & $\delta$ Scuti \\
207504 & W UMa / $\beta$ Lyrae \\
207581 & W UMa \\
208337 & Algol \\
209383 & W UMa \\
209728 & Algol / $\beta$ Lyrae \\
210254 & W UMa \\
211260 & W UMa \\
212142 & W UMa \\
212994 & W UMa \\
213783 & W UMa \\
213786 & W UMa \\
213949 & Algol / $\beta$ Lyrae \\
214088 & W UMa \\
214815 & $\delta$ Scuti \\
215916 & RRab / Cep1O \\
216018 & W UMa \\
216324 & W UMa \\
217014 & W UMa \\
217054 & W UMa \\
218391 & $\delta$ Scuti \\
218440 & W UMa \\
218771 & W UMa \\
219072 & Algol / $\beta$ Lyrae \\
219076 & W UMa \\
219778 & W UMa \\
219999 & W UMa \\
220530 & W UMa \\
220632 & $\beta$ Lyrae / W UMa \\
222540 & W UMa / RRd \\
223171 & W UMa \\
224598 & $\beta$ Lyrae / Algol \\
224783 & W UMa \\
226277 & $\beta$ Lyrae / Algol \\
226796 & W UMa \\
226992 & W UMa \\
227207 & W UMa \\
227744 & W UMa \\
227914 & W UMa \\
228244 & W UMa \\
228293 & W UMa \\
228484 & W UMa \\
228616 & $\delta$ Scuti \\
229158 & W UMa \\
229495 & W UMa / $\beta$ Lyrae \\
229988 & Algol \\
231711 & $\delta$ Scuti \\
231963 & W UMa \\
232401 & Algol \\
233287 & Algol / $\beta$ Lyrae \\
233663 & W UMa \\
233819 & W UMa \\
233822 & W UMa / $\beta$ Lyrae \\
235127 & W UMa \\
235154 & $\beta$ Lyrae / W UMa \\
235353 & RRd / RRab \\
235373 & W UMa \\
235899 & $\delta$ Scuti \\
235941 & W UMa \\
236897 & W UMa \\
237830 & W UMa \\
238138 & W UMa \\
238361 & W UMa \\
238988 & W UMa \\
239777 & W UMa \\
239827 & W UMa \\
240175 & W UMa \\
240177 & $\beta$ Lyrae / W UMa \\
240302 & W UMa \\
240405 & W UMa \\
240562 & W UMa \\
240811 & W UMa \\
241147 & W UMa \\
241466 & W UMa / $\beta$ Lyrae \\
243090 & $\delta$ Scuti \\
243295 & RRab / RRd \\
243520 & W UMa \\
244818 & W UMa \\
244880 & $\delta$ Scuti \\
245466 & W UMa \\
245709 & RRC / RRd \\
246026 & W UMa \\
246036 & W UMa \\
246468 & W UMa / $\delta$ Scuti \\
246476 & W UMa \\
247628 & $\delta$ Scuti \\
247935 & $\delta$ Scuti \\
249084 & W UMa \\
249324 & $\delta$ Scuti \\
249429 & W UMa \\
250817 & Algol \\
250930 & W UMa \\
251418 & Algol \\
251606 & W UMa \\
251710 & W UMa \\
253194 & W UMa \\
254726 & $\delta$ Scuti \\
256025 & $\beta$ Lyrae \\
256298 & W UMa \\
257772 & W UMa \\
258138 & W UMa \\
258518 & RRd / RRab \\
258936 & W UMa \\
259120 & W UMa \\
259166 & W UMa \\
259321 & W UMa \\
259359 & W UMa \\
260224 & $\delta$ Scuti \\
260240 & $\delta$ Scuti \\
260512 & Algol \\
261116 & W UMa \\
262353 & $\delta$ Scuti \\
262375 & W UMa \\
262725 & W UMa \\
263181 & W UMa \\
266182 & W UMa \\
266487 & W UMa \\
266782 & W UMa \\
268375 & $\delta$ Scuti \\
269074 & W UMa \\
269995 & W UMa \\
270233 & W UMa \\
271630 & W UMa \\
272108 & W UMa \\
272656 & W UMa \\
272973 & W UMa \\
273617 & W UMa \\
273738 & W UMa \\
273847 & W UMa \\
274534 & W UMa \\
274965 & W UMa / $\beta$ Lyrae \\
275618 & W UMa \\
276258 & W UMa \\
277014 & W UMa \\
277461 & W UMa \\
277539 & W UMa \\
277992 & W UMa \\
278817 & W UMa \\
279001 & W UMa \\
279436 & $\delta$ Scuti \\
279811 & W UMa \\
280398 & $\beta$ Lyrae / Algol \\
280921 & W UMa \\
282464 & W UMa \\
283579 & $\beta$ Lyrae / Algol \\
284282 & Algol / $\beta$ Lyrae \\
284418 & W UMa \\
284896 & W UMa \\
285403 & $\delta$ Scuti \\
285692 & $\beta$ Lyrae / W UMa \\
285827 & W UMa \\
286273 & W UMa \\
286466 & W UMa \\
286630 & W UMa \\
288432 & Algol / $\beta$ Lyrae \\
289079 & Algol \\
289379 & $\delta$ Scuti \\
290826 & W UMa \\
292142 & $\delta$ Scuti \\
293399 & W UMa \\
293405 & W UMa \\
293627 & Algol \\
293769 & $\delta$ Scuti \\
294525 & W UMa / $\delta$ Scuti \\
296490 & W UMa \\
297270 & W UMa \\
297854 & W UMa \\
298128 & W UMa \\
298817 & W UMa \\
301991 & W UMa \\
303226 & W UMa \\
\end{supertabular}
\end{center}
\vspace{1cm}

\end{appendix}

\end{document}